\documentstyle[11pt,aaspp4,flushrt]{article}
\input{psfig}
\makeatletter

\@ifundefined{chapter}{\def\thebibliography#1{\section*{References\@mkboth
  {REFERENCES}{REFERENCES}}\list
  {\relax}{\setlength{\labelsep}{0em}
        \setlength{\itemindent}{-\bibhang}
        \setlength{\itemsep}{0pt}
        \setlength{\parsep}{0pt}
        \setlength{\leftmargin}{\bibhang}}
    \def\newblock{\hskip .11em plus .33em minus .07em}
    \sloppy\clubpenalty4000\widowpenalty4000
    \sfcode`\.=1000\relax}}%
{\def\thebibliography#1{\chapter*{Bibliography\@mkboth
  {BIBLIOGRAPHY}{BIBLIOGRAPHY}}\list
  {\relax}{\setlength{\labelsep}{0em}
        \setlength{\itemindent}{-\bibhang}
        \setlength{\itemsep}{0pt}
        \setlength{\parsep}{0pt}
        \setlength{\leftmargin}{\bibhang}}
    \def\newblock{\hskip .11em plus .33em minus .07em}
    \sloppy\clubpenalty4000\widowpenalty4000
    \sfcode`\.=1000\relax}}

\newlength{\bibhang}
\let\@internalcite\cite
\def\cite{\let\@citeleft(\let\@citeright)%
    \@ifstar{\citeyear}{\citefull}}
\def\citenp{\let\@citeleft\relax\let\@citeright\relax
    \@ifstar{\citeyear}{\citefull}}
\def\citefull{\def\astroncite##1##2{##1~##2}\@internalcite}
\def\citeyear{\def\astroncite##1##2{##2}\@internalcite}

\def\@citex[#1]#2{\if@filesw\immediate\write\@auxout{\string\citation{#2}}\fi
  \def\@citea{}\@cite{\@for\@citeb:=#2\do
    {\@citea\def\@citea{; }\@ifundefined
       {b@\@citeb}{{\bf ?}\@warning
       {Citation `\@citeb' on page \thepage \space undefined}}%
{\csname b@\@citeb\endcsname}}}{#1}}

\def\@cite#1#2{\@citeleft#1\if@tempswa , #2\fi\@citeright}
\def\@biblabel#1{}

\makeatother

\def\Mpc{{\,h^{-1}\,{\rm Mpc}}}
\def\etal{{\rm et~al.\,}}
\newcommand{\xibar}{\overline{\xi}}

\def\etal{{\rm et~al.\/}}
\def\calD{{\cal{D}}}
\newcommand\approxgt{\mbox{$^{>}\hspace{-0.24cm}_{\sim}$}}
\newcommand\approxlt{\mbox{$^{<}\hspace{-0.24cm}_{\sim}$}}

\begin{document}

\title{Biased-estimations of the Variance and Skewness}
\author{Lam Hui\altaffilmark{1} and
Enrique Gazta\~{n}aga\altaffilmark{2}}
\altaffiltext{1}{NASA/Fermilab Astrophysics Center, Fermi
National Accelerator Laboratory, Batavia, IL 60510; e-mail: \it
lhui@fnal.gov}
\altaffiltext{2}{Consejo Superior de Investigaciones Cient\'{\i}ficas (CSIC),
Institut d'Estudis Espacials de Catalunya, 
Edf. Nexus-104-c/ Gran Capit\'{a} 2-4, 08034, Barcelona,
Spain; e-mail: \it eg@ieec.fcr.es}

\begin{abstract}

Nonlinear combinations of
direct observables are often used to estimate quantities of theoretical
interest. Without sufficient caution, this could lead to biased
estimations. An example of great interest is the
skewness $S_3$ of the galaxy distribution, defined as the ratio of
the third moment $\xibar_3$ and 
the variance squared $\xibar_2^2$ smoothed at some scale $R$.
Suppose one is given unbiased
estimators for $\xibar_3$ and $\xibar_2^2$ respectively, taking a
ratio of the two does not necessarily result in an unbiased estimator
of $S_3$.
Exactly such an estimation-bias (distinguished from the {\it
galaxy}-bias) 
affects most existing
measurements of $S_3$ from galaxy surveys. 
Furthermore, common estimators for $\xibar_3$ and $\xibar_2$ suffer
also from this kind of estimation-bias themselves, because 
of a division by the estimated mean counts-in-cells.
In the case of $\xibar_2$, the bias is equivalent to what is
commonly known as the integral constraint.
We present a unifying treatment allowing all these estimation-biases
to be calculated analytically.
These estimation-biases are in general negative, and
decrease in significance as the survey volume increases, for a
given smoothing scale.
We present a preliminary re-analysis of some existing measurements
of the variance and skewness (from the APM, CfA, SSRS,
IRAS)
and show that most of the well-known systematic discrepancies
between surveys
with similar selection criteria, but different sizes, can 
be attributed to the volume-dependent estimation-biases.
This affects the inference of the galaxy-bias(es) from these surveys.
Our methodology can be adapted 
to measurements of the variance and skewness of, for instance,
the transmission distribution in quasar spectra and
the convergence distribution in weak-lensing maps.
We discuss generalizations to $N > 3$,
suggest methods to reduce the estimation-bias,
and point out other examples in
large scale structure studies which might 
suffer from this type of a nonlinear-estimation-bias.
\end{abstract}

\keywords{cosmology: observations -- large-scale structure of universe
-- methods: statistical} 

\newpage
\section{Introduction}
\label{intro}

There has been a long history of interests, since the pioneering work
of Peebles (\citenp*{peebles80}, \S 18)
in the hierarchical
amplitudes $S_N$, defined as the following ratio:
\begin{equation}
S_N = \xibar_N / \xibar_2^{N-1} \, ,
\label{SNdefKN}
\end{equation}
where $\xibar_N$ is the N-cumulant defined by
$\langle \delta^N \rangle_c$, and $\delta$ is the density
fluctuation smoothed on some scale. 
These quantities are important as a test of the gravitational instability
paradigm (e.g. \citenp{fry84,jusz93,bern94b}),
a probe of possibly non-Gaussian initial conditions 
(e.g. \citenp{silk91,gm96,gf97})
as well as a measure of the galaxy-bias (e.g.  \citenp{fg93,fg94b}).
However, it has also been a puzzle for quite some time that
different galaxy surveys yield discordant values for $S_N$
(e.g. Table \ref{s3s4}, for $N = 3,4$). 
While some of the differences no doubt arise from the fact that
galaxies selected in different ways might have different
galaxy-biases, not all of the differences can be convincingly
explained away in such a manner. For instance, a comparison between
optically selected galaxy-catalogs (in Table \ref{s3s4}) reveal a substantial
and systematic difference
between the measured values of $S_N$: redshift surveys 
consistently yield lower values compared to the larger angular catalogues
(e.g. compare APM/LICK/EDSGC values with those from CfA/SSRS
in Table 1; note that the IRAS galaxies are infrared-selected; note also
an exception to this rule in the measurements by \citenp{kimstrauss98}).
A rather large relative galaxy-bias between these two sets of
catalogues would have to be invoked to reconcile them.

Three alternative explanations are possible. 
The first is that redshift space distortions
tend to suppress $S_N$, but it has been shown
to be not sufficient to explain the systematic differences, 
especially on large scales
(\citenp{fg94}). The second is that the local volume 
(sampled by the redshift surveys) just happens to have 
a smaller $S_N$ compared to the true $S_N$ which is presumably
measured in the larger angular surveys i.e. the local universe
is not a fair sample (e.g. \citenp{g94}).
 This is related to the question of
the homogeneity scale of our universe which has been a subject
of some debate (see e.g. \citenp{wu98}
). The third is to
blame it on the 
estimator for $S_N$:  
it yields a value that is on the average biased low, with
the bias (distinguish this from the {\it galaxy}-bias) getting worse
as the survey becomes smaller. 
We will demonstrate that the third contributes to a significant
fraction of the systematic differences between the different
measurements. While a thorough analysis detailing exactly how much
each of these factors contribute is beyond the scope of this paper,
we can safely conclude that 
inference of large sampling fluctuations, or a large
relative galaxy-bias, based on measurements of $S_3$, are
unwarranted.

A very closely related estimation-bias  has been discussed before by
Colombi et al. 
(\citenp*{cbs94})
as a finite-volume effect.\footnote{Related
ideas have also been considered by Bromley 1998, private
communication.} They attributed this to the abrupt 
cut-off of the count probability at some finite number of particles
because of the finite size of a survey. They proposed a way to correct
for this bias by extending the tail of the count probability using a
few phenomenological parameters calibrated from simulations
(see also \citenp{fg94,mbms97}). The method works reasonably well
for small smoothing scales, but not for large ones, mainly because the
probability-distribution-tail becomes noisy in the latter case 
(Colombi 1998, private communication). Moreover, the method
is feasible only with a dense-sampled survey (\citenp{bouchet93}).
More recently, Szapudi \& Colombi (\citenp*{sc96}) and
Colombi et al. \cite*{css98} discussed how a finite survey-volume affects
the errors (i.e. the variance around the mean) in the estimators of 
the N-cumulants $\xibar_N$ (or the related   
factorial moments),
but not the mean or bias of the estimator for $S_N$. 
 
However, there has been no attempt to explain quantitatively
the differences in the measured $S_N$ from different surveys of
similar selection criteria in terms
of the finite-volume effect. This is, perhaps, in part due to the lack
of an analytical 
estimate of the systematic bias of the standard estimators for $S_N$.
As we will show, a remarkably simple statistical fact allows
just such a calculation to be done, while
clarifying the origin of this bias, and relating it to other known
biases in large scale structure statistics, such as the integral
constraint.

Consider the following elementary statement:
\begin{equation}
\langle {\hat A \over \hat B} \rangle \ne {\langle \hat A \rangle 
\over {\langle \hat B \rangle}}
\label{AB}
\end{equation}
where $\langle \, \rangle$ denotes ensemble averaging, and $\hat A$
and $\hat B$ are two random variables or estimators. This statement holds
generically, except in special cases such as when $\hat A$ and $\hat
B$ are constants.

The standard method of estimating $S_N$ is to form estimates
of $\xibar_N$ and $\xibar_2$ separately, and then take an appropriate
ratio of the two. However, even in the ideal case
where one has an unbiased estimator of the numerator ($\xibar_N$; let
us call this estimator $\hat A$, i.e. $\langle \hat A \rangle = \xibar_N$)
and an unbiased estimator of the denominator ($\xibar_2^{N-1}$; let us call
the estimator $\hat B$, i.e. $\langle \hat B \rangle = \xibar_2^{N-1}$)
\footnote{We will show shortly that even this ideal case does
not hold in reality.}, taking a ratio of the two estimators
does not necessarily result in an unbiased estimator of $S_N$.
This is captured by the statistical statement in eq. (\ref{AB}). 
We will refer to this kind of estimation-bias as the ratio-bias.

More generally, nonlinear combinations of unbiased estimators
should be treated with great care. For instance, suppose ${\hat{\xibar}}_2$ is
an unbiased estimator of $\xibar_2$ such that $\langle {\hat{\xibar}}_2
\rangle = \xibar_2$. It is virtually guaranteed that $\langle
({\hat{\xibar}}_2)^2 
\rangle \ne (\xibar_2)^2$. We will refer to this kind of bias
generally as a nonlinear-estimation-bias, of which the ratio-bias is a
particularly simple and common example. 
In this paper, we will sometimes be abusing the terminology by using
the two terms interchangeably.

The well-known integral constraint, in the case of measurements
of the two-point function, can in fact be understood
as a ratio bias. The two-point function $\xi_2 (i,j)$, where $i$ and $j$
are two cells separated by some distance,
is by definition $\langle \delta_i \delta_j \rangle$ where 
$\delta_i$ is the overdensity at cell $i$.
The catch is that one directly observes only $n_i$, the number
of particles/galaxies in a cell. An estimate of the mean number
count $\bar n$ has to be made in order to assign a value $\delta_i$
for each cell. This is generally taken to be $\sum_i n_i / N_T$ where
$N_T$ is the total number of cells, let us call this estimator ${\hat
{\bar n}}$.  
The problem is, of course, that $\langle (n_i - {\hat {\bar n}})
(n_j - {\hat {\bar n}}) / {\hat {\bar n}}^2 \rangle \ne \xi (i,j)$,
because of the estimator $\hat {\bar n}$ in the denominator. As we
will see, this estimation-bias is in general negative.
This is what the integral constraint is about: that the measured
two-point function 
is generally biased low because the
mean number density is estimated from the same survey from which
$\xi_2$ is being measured. This turns out to be the ratio-bias in
disguise. It is also easy to see that estimates of $\xi_N$ would suffer
from a similar bias.

Peebles \cite*{peebles80} first pointed out the integral-constraint-bias
for estimating $\xi_2$, but his treatment only gave the large
scale estimation-bias. 
Bernstein \cite*{b94}, building on earlier work by Landy \& Szalay
\cite*{landy93}, developed a perturbative approach (not perturbative
in the usual sense of small density fluctuations, but perturbative
in the small quantity: the average of the two-point function over 
the volume of the survey) to
compute the full integral-constraint-bias for $\xi_2$, which accounted
for the small-scale bias as well. (See also \citenp{kerscher98} for a
related recent discussion.) Obviously, 
the integral-constraint bias affects also measurements of
the one-point analogue, or the volume-average, of $\xi_2$ i.e. the variance
$\xibar_2$. This integral-constraint-bias generally decreases in magnitude
with increasing survey size, causing measurements of the correlation length
from $\xi_2$ or $\xibar_2$ to increase with sample depth, an
effect that has been observed before (\citenp{davis88,bouchet93}).
Hence, in the case of the two-point function or its volume average,
the finite-volume-effect pointed out by Colombi et al.
\cite*{cbs94} is none other than the integral-constraint-bias.

Adopting the techniques of Bernstein \cite*{b94}, we 
compute analytically the biases of the standard estimators for
$\xibar_N$ and $S_N$, for $N= 2, 3$. 
The methodology for a general $N$ is presented in \S \ref{derivation}.
For simplicity, we illustrate how to keep track of the
perturbative-ordering by going into details of the calculation for $N
= 2, 3$. These cases are also of special interest because 
many measurements of $\xibar_2$, $\xibar_3$ and $S_3$ exist in the literature.
We go over the calculation of the estimation-biases for these
three quantities in \S \ref{egN3}. 
For readers not interested in the details: much of the section can be
skipped; the main results are in eq. (\ref{bK2PT}), (\ref{bK3PT}) \&
(\ref{biasS3noshotPT}). 

We next check in \S \ref{numerical} our analytical results using
N-body simulations  
of the SCDM (Standard Cold-Dark-Matter) and LCDM (Lambda or
Low-Density Cold-Dark-Matter) models. 
Ensemble averages of the standard estimators for $\xibar_2$ and $S_3$
are computed 
by using $10$ realizations for each chosen 
model and for various sample-volumes. The overall agreement is
excellent. We also 
introduce a way to correct for the analytical estimates when the
estimation-bias becomes so large that the perturbative approach
in \S \ref{egN3} breaks down.

In \S \ref{catalogue}, we present a first step towards a re-evaluation
of existing measurements of the variance and skewness from the
CfA/SSRS/APM surveys. We study simulated
CfA/SSRS/APM catalogues 
with the appropriate sizes, which include the effects
of redshift distortions as well as sparse-sampling.
We then consider a preliminary correction of 
the existing measurements of the variance and skewness from
these surveys, based on our findings in \S \ref{egN3} and \S \ref{numerical}.
The correction is necessarily model-dependent
(dependent on the power spectrum and the galaxy-biasing
assumed), but
it appears that most of the systematic differences between
these surveys can be explained by the estimation-biases, 
under reasonable assumptions.
A thorough analysis of the remaining differences
would require a careful study of, among other things, projection
effects and redshift distortions. We leave this
for future work.

Finally, we conclude in \S \ref{conclude} with a discussion of
methods for measuring $S_N$ that might be subject to a less severe
estimation-bias. 
We list other large scale structure statistics which might
also suffer from this type of a nonlinear-estimation-bias.
We also discuss applications of our findings outside conventional
galaxy surveys, such as the Lyman-alpha forest, high-redshift
Lyman-break galaxy surveys and weak-lensing maps.

\section{Biases of the Standard Estimators for $\xibar_N$ and $S_N$}
\label{derivation}

\subsection{Definitions}
\label{notation}

The standard estimator for $S_N$ is given by:
\begin{equation} 
\hat{S}_N = \hat{\xibar}_N / ({\hat{\xibar}_{2}})^{N-1} \, .
\label{hatSN}
\end{equation}

We use $\hat{}$ to denote estimators of quantities we are interested in.

$\hat{\xibar}_2$ is the esimator for the variance:
\begin{equation}
\hat{\xibar}_2 = {1 \over N_T} \sum_i (\hat\delta_i)^2 - 
\hat{\xibar}_2^{\rm shot} .
\label{K2hat}
\end{equation}
Imagine that the survey is divided into many very small cells,
so small that the number of particles/galaxies in each cell is either 1 or 0. 
The index $i$ above denotes such a cell, and $N_T$ is the total number
of such cells. $\hat\delta_i$ is an estimate of the local overdensity
smoothed over some given radius $R$. We will assume top-hat
smoothing in this paper. In other words:
\begin{equation}
\hat\delta_i = \sum_j {n_j - \hat{\bar n} \over \hat{\bar n}} W_T (i,j)
\, ,
\label{deltahat}
\end{equation}
where $n_j$ is equal to 1 if there is a galaxy and 0 otherwise,
$\hat{\bar n}$ is an estimate of the mean density of the survey, and
$W_T (i,j)$ is the top-hat smoothing 
window. The estimator $\hat{\bar n}$ is
\begin{equation}
\hat{\bar n} = {1\over N_T} \sum_i \sum_j n_j W_T(i,j) \, .
\label{hatbarn}
\end{equation}

We have not stated explicitly how to handle edge-effects.
For instance, what should be done when the top-hat window $W_T$ 
overlaps with the boundary in eq. (\ref{deltahat})? 
Note that $\sum_j W_T(i,j) = 1$ only 
for $i$ sufficiently far away from edges.
If one adopts the strategy that one picks only $i$'s in eq.
(\ref{deltahat}) such that the top-hat does not overlap with the
boundary, then the corresponding $N_T$ in eq. (\ref{K2hat}) should not
be the total number of all infinitesimal cells in the sample, but
a smaller number: the number of centers (at the infinitesimal cells) 
of top-hats whose windows do not overlap with the boundary. 
The $j$-index in eq. (\ref{deltahat}) should, on the other hand, 
range over the whole survey volume, up to the boundary.

The standard shot-noise correction is given by 
\begin{equation}
\hat{\xibar}_2^{\rm shot} = {1 \over \hat{N_R}}
\label{K2shot}
\end{equation}
where $\hat{N_R}$ is the estimated mean number of particles within
a smoothing top-hat of size $R$, in other words, it is
$\hat{\bar n} V_R$ where $V_R$ is the volume of the top-hat.

Phrased in the above manner, the estimator $\hat{\xibar}_2$ is
equivalent to the standard estimator for the variance using the
counts-in-cells method, 
infinitely sampled (see \citenp{s98}). 

The estimator for the N-th cumulant is defined similarly:
\begin{equation}
\hat{\xibar}_N = {1 \over N_T} \sum_i (\hat\delta_i)_c^N - \hat{\xibar}_N^{\rm shot} \, 
\label{KNhat}
\end{equation}
where the subscript $c$ denotes the connected 
part of the sum, and $\hat{\xibar}_N^{\rm shot}$ is a shot-noise
correction generalizing $\hat{\xibar}_2^{\rm shot}$.
It is worthwhile at this point to introduce the continuum
notation. We will replace $\sum_i/ N_T$ by $\int dV_i / V_T$,
where $dV_i$ is the volume of the i-th cell and $V_T$ is the total
volume,  
and $W_T (i,j)$ by the continuum top-hat $W (i,j)$ normalized so that
$\int dV_j \, W(i,j) = 1$ for $i$ sufficiently far away from edges.

It is also worthwhile to note that factorial moments
are sometimes used to estimate $\xibar_N$, which represents
a convenient way to eliminate the shot-noise contribution
(\citenp{ss93}), but otherwise results in
the same estimator for $\xibar_N$ as in eq. (\ref{KNhat}).

\subsection{Derivation of the Estimation-Biases: an Outline}
\label{outline}

The integral-constraint-bias for $\hat{\xibar}_2$ arises from the fact
that the true mean density $\bar n$ is unknown, and has to estimated from the
same sample from which one tries to measure the variance.
To derive it, 
let us first write the estimator for the mean density as $\hat{\bar n} = \bar
n (1 + \alpha)$, where $\hat{\bar n}$ is given 
in eq. (\ref{hatbarn}), and $\alpha$ is a small fluctuation from the
true mean. Then one can express the estimator for the overdensity
$\hat\delta_i$ (eq. 
[\ref{deltahat}]) as
\begin{equation}
\hat\delta_i = (\delta_i - \alpha) (1 -\alpha + \alpha^2 - ...) \, ,
\label{hatdeltaexpand}
\end{equation}
where $\delta_i$ is now the true overdensity (smoothed with the same
top-hat as $\hat\delta_i$ is) i.e. $\delta_i$ is $n_i / \bar
n - 1$, appropriately smoothed. $\alpha$ is by definition equal to
\begin{equation}
\alpha = {1\over V_T} {\int dV_i \delta_i} \, .
\label{alpha}
\end{equation}

Substituting the above into the expression for $\hat{\xibar}_2$ (eq.
[\ref{K2hat}]), one can expand in $\alpha$ and write down the 
ensemble average of $\hat{\xibar}_2$ order by order. It can be easily
shown that the zero-th order term (no $\alpha$) of $\langle
\hat{\xibar}_2 \rangle$ gives $\xibar_2$, the true variance. 
The rest of the terms represent the integral-constraint-bias.
A key question
is by what order one should stop, which we will discuss in detail
in \S \ref{K2}. It suffices to note here that, strictly speaking,
$\alpha$ by itself cannot be used as an ordering-parameter, because it
is a random variable (depends on the data) which gets
ensemble-averaged.
Let us denote the integral-constraint-bias for
$\hat{\xibar}_2$, or more generally, $\hat{\xibar}_N$ by:
\begin{equation}
\langle \hat{\xibar}_N \rangle \equiv \xibar_N (1 + {\Delta_{\xibar_N}
\over \xibar_N}) \, . 
\label{biashatKN}
\end{equation}
As we will see, the fractional bias ${\Delta_{\xibar_N} / \xibar_N}$ 
becomes small for a large enough survey. This is the limit in which we
will be working. How large is large, or how small is small, is the
question we would like to address. 

It is easy to see that the above methodology can be adopted for
computing the ensemble average of the standard estimator
for the hierarchical amplitude,
$\langle \hat S_N \rangle$. The key idea
is to assume the denominator part of the estimator fluctuates about
its mean, and expand in that fluctuation. In other words, for $S_N$,
let us first assume
\begin{equation}
\hat{\xibar}_2 \equiv \xibar_2 (1 + {\Delta_{\xibar_2} \over \xibar_2}) ( 1 + \epsilon) \, ,
\label{epsilon}
\end{equation}
where $\epsilon$ is a small fluctuation of the measured $\hat{\xibar}_2$
from its mean (which is offset from the true $\xibar_2$ by a bias; eq.
[\ref{biashatKN}]). Note that $\epsilon$ depends on the data, and
so cannot be taken out of $\langle \rangle$.

Putting eq. (\ref{biashatKN}) and (\ref{epsilon})
into eq. (\ref{hatSN}), one obtains the following expression for
the mean of $\hat S_N$:
\begin{eqnarray}
\label{biasSN}
\langle \hat S_N \rangle &=& S_N (1 + {\Delta_{S_N} \over S_N}) \\ 
\label{biasSN2}
{\Delta_{S_N} \over S_N} &=& {\Delta_{\xibar_N} \over \xibar_N} - (N-1)
{\Delta_{\xibar_2} \over \xibar_2} - (N-1) {\langle \epsilon \hat{\xibar}_N \rangle  \over
\xibar_N} + (N-1){N\over 2} {\langle \epsilon^2 \hat{\xibar}_N \rangle  \over
\xibar_N} + ... \, .
\end{eqnarray}
Note that all terms up to $\epsilon^2$ are displayed above, except
for the terms: $- (N-1) (\Delta_{\xibar_2} /\xibar_2) [\langle
\epsilon \hat{\xibar}_N \rangle  /
\xibar_N]$ and $ (N-1){N\over 2} (\Delta_{\xibar_2} /\xibar_2)
[\langle \epsilon^2 \hat{\xibar}_N \rangle  / \xibar_N]$.
As we will show later on, the terms $\Delta_{\xibar_2} /\xibar_2$,
$\langle \epsilon \hat{\xibar}_N \rangle / \xibar_N$ and 
$\langle \epsilon^2 \hat{\xibar}_N \rangle  / \xibar_N$ are
all of the order of a small parameter in which we will be expanding:
hence the dropping of products involving them.

It is instructive to divide the net bias of the estimator $\hat
S_N$ into two different contributions. One is due to
the integral-constraint-bias discussed above: namely that both
${{\hat{\xibar}}}_N$ in the numerator and ${{\hat{\xibar}}}_2$ in the denominator
(see eq. [\ref{hatSN}]) are biased estimators. This gives the first two
terms on the right hand side of eq. (\ref{biasSN2}). The other
contribution arises from the fact that the estimator
$\hat S_N$ is the ratio of two other estimators (to be accurate, it
is in fact some nonlinear combinations of ${{\hat{\xibar}}}_N$
and ${{\hat{\xibar}}}_2$). This is the rest of
the terms in eq. (\ref{biasSN2}). Let us give these two kinds of terms
explicit names:
\begin{eqnarray}
\label{intconstbias}
&& {\Delta_{S_N}^{\rm int. constr.} \over S_N} = {\Delta_{\xibar_N}
\over \xibar_N} - (N-1) 
{\Delta_{\xibar_2} \over \xibar_2} \, , \\
\label{ratiobias}
&& {\Delta_{S_N}^{\rm ratio} \over S_N} = - (N-1) {\langle \epsilon
\hat{\xibar}_N 
\rangle  \over  
\xibar_N} + (N-1){N\over 2} {\langle \epsilon^2 \hat{\xibar}_N \rangle  \over
\xibar_N} \, ,
\end{eqnarray}
the integral-constraint-bias and the ratio-bias of $\hat S_N$
respectively. We are slightly abusing the terminology here because
the integral-constraint-bias is of course itself a form of a
ratio-bias.
As we will show below, it turns out that the terms contributing to the
integral-constraint-bias
partially cancel each other, leaving the ratio-bias
to be the dominant contribution to the net bias of $\hat S_N$ on large
scales.

Note that we have implicitly assumed ${\Delta_{\xibar_N} / \xibar_N}$, $(N-1)
{\Delta_{\xibar_2} / \xibar_2}$ and $(N-1) {\langle \hat{\xibar}_N \epsilon \rangle /
\xibar_N}$ and $(N-1)N \langle \epsilon^2 \hat{\xibar}_N \rangle /
2 \xibar_N$ are all small, and that terms we have ignored in eq.
(\ref{biasSN2}) are somehow all of higher order.
What is the correct ordering parameter in which we are expanding?
Again, the quantity $\epsilon$ cannot be used directly to
keep track of the ordering, because it is data-dependent. For
instance, after taking
the expectation values, it could happen that terms that contain
$\epsilon^2$ are actually comparable to terms linear in $\epsilon$.
We will see that this is indeed the case in the next section.

eq. (\ref{biasSN}) implies
the bias of $\hat S_N$ depends in general on the $M$-point correlation
functions. The strategy we adopt in this paper is to 
assume the following hierarchical relation: $\xi_M \sim \xi_2^{M-1}$, which is
motivated by perturbation theory but has been observed to hold in the 
highly nonlinear regime as well. We do not need to assume anything, however, 
about the configuration or scale dependence (or independence) of the
hierarchical 
amplitudes. Using this relation, it can be shown that 
(we will demonstrate this explicitly for $S_3$) the terms we
have kept in $\Delta_{S_N} / S_N$ (eq. [\ref{biasSN2}]) all contain
terms linear in the following quantity:
\begin{equation}
\xibar_2^L \equiv {1 \over V_T^2} \int dV_i dV_j \xi_2 (i,j)
\label{K2L}
\end{equation}
where $\xi_2$ is a smoothed version of the 2-point function defined as
\begin{equation}
\xi_2 (i,j) = \int dV_k dV_l \, \, \xi_2^{\rm usmth.}(k,l) \, W(k,i) W(l,j)
\label {xi2R}
\end{equation}
where $W$ is the top-hat smoothing window of some radius $R$, and
$\xi_2^{\rm usmth.}$
is the unsmoothed 2-point function. 
$\xibar_2^L$ is the 2-point function averaged over the whole survey (of
size $L$). For a survey to be of any use at all,
this quantity has to be much smaller than 1. Because of the relatively large 
coefficients that will be multiplying it in $\Delta_{S_N} /S_N$, as will be shown 
below, the fractional bias in $\hat S_N$ could be
non-negligible, even for a relatively large survey volume.  

One should therefore view the derivation of the bias in $\hat S_N$ in
this paper as an expansion in the small parameter $\xibar_2^L$. 
As we will see, the same applies to $\hat{\xibar}_N$.
Indeed, it can be shown that all the terms we have kept in eq.
(\ref{biasSN}) contain terms linear in $\xibar_2^L$, and terms 
we have ignored are all of higher order ($[\xibar_2^L]^2$, etc).
We will demonstrate the reasoning with an example: $N = 3$.

\section{Estimation-Biases for the Variance, the Third Moment and the Skewness}
\label{egN3}

\subsection{The Integral-Constraint-Bias for $\hat{\xibar}_2$}
\label{K2}

The integral-constraint-bias for the two-point function has been known
for a long time (see e.g. \citenp{peebles80,b94,tegmark97}). Our
treatment follows most closely that of Bernstein \cite*{b94}, and the
emphasis is on techniques that can be generalized to
${\hat{\xibar}}_3$ and $\hat S_3$.

Following the strategy outlined in \S \ref{outline}, namely combining
eq. (\ref{hatdeltaexpand}), (\ref{alpha}) with (\ref{K2hat}), 
we obtain
\begin{eqnarray}
\label{bK2}
\langle \hat{\xibar}_2 \rangle = && {1\over V_T} \int dV_i \langle
\delta_i^2 \rangle - \langle \hat{\xibar}_2^{\rm shot} \rangle \\
\nonumber
&& - {1\over V_T^2} \int dV_i dV_j \langle \delta_i \delta_j \rangle -
{2\over V_T^2} \int dV_i dV_j \langle \delta_i^2 \delta_j \rangle \\
\nonumber 
&& + {3\over V_T^3} \int dV_i dV_j dV_k \langle \delta_i^2 \delta_j
\delta_k \rangle + {4\over V_T^3} \int dV_i dV_j dV_k \langle \delta_i
\delta_j \delta_k \rangle + ... 
\end{eqnarray}
where we display all terms up to $O(\alpha^2)$ (see eq. [\ref{alpha}]).

As we have explained before, $\alpha$ cannot really be used as an
ordering-parameter, because it is a random variable which gets
averaged over. The key to the above expansion is instead to keep only
terms up to linear order in $\xibar_2^L$ (eq. [\ref{K2L}]). 

Ignoring shot-noise for now, the first term on the right gives
us the true variance $\xibar_2$. This is the zero-th order term.
The rest of the terms represent the integral-constraint-bias for
${\hat{\xibar}}_2$, and are all of order $\xibar_2^L$ or higher.
Let us check.

1. The first term on the second line of eq. (\ref{bK2}) is none other
than $-\xibar_2^L$ itself (eq. [\ref{K2L}]). 

2. The next term, again ignoring
shot-noise for now, gives $(-2/V_T^2) \int dV_i dV_j \xi_3 (i,i,j)$. 
Applying the hierarchical relation $\xi_M \sim \xi_2^{M-1}$, i.e.
$\xi_3 (i,i,j) \sim \xi_2 (i,i) \xi_2 (i,j) + ...$, we can see that
the term $\xi_2 
(i,i) \xi_2 (i,j) = \bar\xi_2 \xi_2 (i,j)$, when integrated over $i$
and $j$, will give rise to a term proportional to $\xi_2^L$. Hence,
the term $(-2/V_T^2) \int dV_i dV_j \xi_3 (i,i,j)$ contains
a linear piece and should not be thrown away. Note
that we need not make any assumption about the configuration or
scale dependence of the hierarchical amplitudes. In other words,
we should keep the term $(-2/V_T^2) \int dV_i dV_j \xi_3 (i,i,j)$ as is,
rather than as, say $(-2/V_T^2) \bar\xi_2 \int dV_i dV_j \xi_2 (i,j)$.
The hierarchical
relation is used strictly for keeping track of the
ordering at this stage.

3. The integrand in the next term, 3 $\langle \delta_i^2 \delta_j \delta_k
\rangle$, can be broken up into several disconnected pieces:
$3 \langle \delta_i^2 \delta_j \delta_k \rangle = 3 [\xi_2 (i,i) \xi_2
(j,k) + 2 \xi_2 (i,j) \xi_2 (i,k) + \xi_4 (i,i,j,k)]$. 
It is easy to see that the second piece gives something that is second
order in $\xibar_2^L$, when integrated over, and so does the $\xi_4
(i,i,j,k)$ piece, assuming again the hierarchical relation.
The only piece that survives is then, after integration, $3 \xibar_2
\xibar_2^L$. 

4. Lastly, the $(4/V_T^3) \int dV_i dV_j dV_k \langle
\delta_i \delta_j \delta_k \rangle$ term in eq. (\ref{bK2}) is second
order in $\xibar_2^L$, again by applying the hierarchical relation.

More generally, it can be seen that all terms, including those not explicitly
displayed, in the expansion in eq. (\ref{bK2}) are of the form
$V_T^{-m} \int dV_{i_1} ...
dV_{i_m} \langle \delta_{i_1}^\gamma \delta_{i_2} ... \delta_{i_m}
\rangle$, where $\gamma$ is 1 or 2.
Our arguments above show that only terms with $m = 1$, or $m = 2$, or
$m = 3$ and $\gamma = 2$, contain pieces linear in $\xibar_2^L$. 
The reader can convince himself or herself that all other terms
are of higher order.

Putting everything together, ignoring shot-noise, and using the
definition of the fractional bias in eq. (\ref{biashatKN}), we obtain
to linear order in $\xibar_2^L$,
\begin{eqnarray}
\label{bK22}
{\Delta_{\xibar_2} \over \xibar_2} = && - {1\over \xibar_2 V_T^2} \int
dV_i dV_j \xi_2 
(i,j) - {2\over \xibar_2 V_T^2} \int
dV_i dV_j \xi_3 (i,i,j) \\ \nonumber
&& + {3 \over V_T^2} \int dV_i dV_j \xi_2 (i,j)  \, .
\end{eqnarray}
The above result is consistent with that of Bernstein \cite*{b94} for
the integral-constraint-bias of the two-point function.
The first term on the right was obtained by Peebles \cite*{peebles80}
via a different argument.

How about shot-noise? A term like $\langle \delta_i \delta_j \rangle$
that shows up in eq. (\ref{bK2})  
includes both the cosmic 2-point correlation $\xi_2 (i,j)$ (see eq.
[\ref{xi2R}]) and a shot-noise contribution due to Poisson sampling
(see e.g. \citenp{fkp94}):
\begin{equation}
\langle \delta_i \delta_j \rangle = \xi_2 (i,j) + {1\over \bar n} \int
dV_k W(k,i) W(k,j) 
\label{shotnoise}
\end{equation}
It can be shown that all shot-noise
contributions to $\Delta_{\xibar_2} / \xibar_2$ can
be expanded in either $1/N_R$ or $1/N_L$ where $N_R$ is the mean
number of particles in a cell of size $R$, and $N_L$ is the mean
number of particles in the whole survey.
For instance, the term $V_T^{-1} \int dV_i \langle \delta_i^2 \rangle$
has a Poisson term $1/N_R$, if one makes use of eq. (\ref{shotnoise}),
together with the fact that $W$ is a top-hat of size $R$ with the
normalization $\int dV_j W(i,j) = 1$. This gets canceled by
part of the shot-noise correction term $- \langle {\hat{\xibar}}_2^{\rm
shot} \rangle$ (eq. [\ref{bK2}]). On the other hand,
a term like $V_T^{-2} \int dV_i dV_j \langle \delta_i \delta_j
\rangle$ gives us a Poisson term of the order of $1/ N_L$.
In general, $1/ N_L$ is a very small quantity, and so we can ignore
all terms of order $1/N_L$. There are, for instance, Poisson pieces
 linear in $\xibar_2^L$ in the $\langle \delta_i \delta_j \delta_k \rangle$
term in eq. (\ref{bK2}), but they are also of order $1/N_L$, and so
can be ignored.

How about the $O (1/N_R)$ terms in eq. (\ref{bK2})? 
Without going into details, it can be shown that
the only $O (1/N_R)$ contributions are: 
$- \xibar_2^L / N_R$ from the $- \langle {\hat{\xibar}}_2^{\rm
shot} \rangle$ term, $- 2 \xibar_2^L / N_R$ from the fourth term on the
right, and $ 3 \xibar_2^L / N_R$ from the fifth term on the right.
Therefore,
to $O (1/N_R)$, the shot-noise terms miraculously cancel!
It is interesting to note that this cancellation is possible only
because the integral-constraint-bias in the shot-noise correction itself
is taken into account i.e. $\langle {\hat{\xibar}}_2^{\rm
shot} \rangle = \langle 1/ \hat N_R \rangle = (1 + \langle \alpha^2 \rangle
...)/N_R$  (see eq. 
[\ref{K2shot}] \& [\ref{alpha}]).

We will see in \S \ref{numerical} that our estimations
of the biases in the variance and skewness which ignores
shot-noise are in fact quite 
accurate. With no further justification, we will ignore
shot-noise terms in the rest of our derivation, which
substantially simplifies our expressions. The cancellation to $O (1/N_R)$ 
here for $\Delta_{\xibar_2} / \xibar_2$ can be taken as suggestive evidence
that shot-noise is unimportant for the estimation-biases we are
interested in;
we will rely on the numerical simulations in \S \ref{numerical}
for further support. We should emphasize, however, we are not saying that
there is no need to subtract out shot-noise when estimating the
variance and skewness themselves.

Finally, how about edge-effects? It is worth emphasizing that no
assumptions about the edge-effects being small need to be made in
deriving eq. (\ref{bK22}). One only has to be careful about
volume over which the integration is done and what $V_T$ has to be.
As we have mentioned earlier in \S \ref{notation}, one way to deal
with the boundary is to use only cells that do not overlap with
the edges. In that case, the dummies of integration $i$ and $j$
should range over the inner part of the survey, where
any cell that is centered within it does not cut into the boundary.
Similarly, $V_T$ should be chosen to be the volume of that inner
region. We will discuss how to approximate such integrals in
\S \ref{everything}.

\subsection{The Integral-Constraint-Bias for $\hat{\xibar}_3$}
\label{K3}

Similarly, one can derive the integral-constraint-bias for 
the third cumulant using eq. (\ref{KNhat}),
(\ref{biashatKN}) and 
(\ref{hatdeltaexpand}) . Ignoring shot-noise, and again,
 keeping only
terms to first order in $\xibar_2^L$, 
the integral-constraint-bias for $\xibar_3$ is given by
\begin{eqnarray}
\label{bK3}
{\Delta_{\xibar_3} \over \xibar_3} =&& -{3\over \xibar_3 V_T^2}
\int dV_i dV_j \xi_3 (i,i,j) - {9 \xibar_2 \over \xibar_3 V_T^2} \int dV_i dV_j
\xi_2 (i,j)  \\ \nonumber 
&& - {3 \over \xibar_3 V_T^2}
\int dV_i dV_j \xi_4 (i,i,i,j)
+ {9 \xibar_2 \over \xibar_3 V_T^2} \int dV_j dV_k \xi_2 (j,k) \\ \nonumber
&& + {6\over
V_T^2} \int dV_j dV_k \xi_2 (j,k)
\end{eqnarray}
Note that by essentially the same reasoning as in the case of 
$\hat{\xibar}_2$, one only needs to consider terms up to $\alpha^2$ in
deriving the above.

\subsection{The Estimation-Bias $\hat S_3$}
\label{ratiobiasS3}

The integral-constraint-bias for $\hat S_3$ can be simply read off
from eq. (\ref{intconstbias}), (\ref{bK22}) \& (\ref{bK3}). 
The ratio-bias for $\hat S_3$, on the other hand, follows from eq.
(\ref{ratiobias}). Substituting the definition of $\epsilon$ from eq.
(\ref{epsilon}), we obtain
\begin{eqnarray}
\label{K3epsilon}
{\Delta_{S_3}^{\rm ratio} \over S_3} 
=&&  {-2 \over \xibar_3 \xibar_2 (1 +
{\Delta_{\xibar_2}\over \xibar_2})} \langle \hat{\xibar}_3 [\hat{\xibar}_2 - \xibar_2 (1 + {\Delta_{\xibar_2}\over
\xibar_2})] \rangle \\ \nonumber 
&& + { 3\over \xibar_3 \xibar_2^2 (1 +
{\Delta_{\xibar_2}\over \xibar_2})^2 } \langle \hat{\xibar}_3 [\hat{\xibar}_2 - \xibar_2 (1 + {\Delta_{\xibar_2}\over
\xibar_2})]^2 \rangle\, . 
\end{eqnarray}

Substituting eq. (\ref{bK2}), (\ref{bK3}) and (\ref{K3epsilon}) into
eq. (\ref{biasSN}) for $N=3$, ignoring shot-noise terms and keeping
only terms linear in $\xibar_2^L$, it can
be shown that
\begin{eqnarray}
\label{bS3epsilon}
{\Delta_{S_3}^{\rm ratio} \over S_3} =&& 
- {6\over V_T^2 \xibar_3} \int dV_i dV_j \xi_3 (i,j,j) - {2\over V_T^2 \xibar_2
\xibar_3} \int dV_i dV_j \xi_5 (i,i,i,j,j) \\ \nonumber 
&& + {3\over V_T^2 \xibar_2^2} \int
dV_j dV_k \xi_4 (j,j,k,k) \, ,
\end{eqnarray}
where $\xi_N$ is the connected N-point function.
The first two terms here arise from the first term on the right hand
side of eq. (\ref{K3epsilon}), and the last term from the second one.

The reasoning used to arrive at the above expression is again
very similar to the case of $\hat{\xibar}_2$. But there are a few new
tips to keep in mind. 

1. Consider a term like $\langle \delta_i^3
[\delta_j^2 - \xibar_2 (1 + \Delta_{\xibar_2}/\xibar_2)] \rangle$,
which arises from the first term on the right of eq.
(\ref{K3epsilon}). It can be seen that a disconnected piece with the
$j$ index all by itself, such as $\langle \delta_i^3 \rangle \langle
\delta_j^2 \rangle$, is going to be canceled, by $-\langle \delta_i^3
\rangle \xibar_2 
(1+\Delta_{\xibar_2}/\xibar_2)$. More generally, it can be seen that
all terms in the expansion of eq. (\ref{K3epsilon}) contain
an integrand of the form 
$\langle \delta_i^3 \delta_{j_1}^2 \delta_{j_2}^2 ... \delta_{j_m}^2 \rangle$,
any disconnected piece of which with the $j_1$, or ... $j_m$ index all by
itself is going to get canceled. In other words, the $j$-indices must
be connected to each other, or to $i$.

2. As before, integrals over products of the two-point function, of
the form $\langle \delta_i \delta_j \rangle \langle \delta_j \delta_k
\rangle$ for instance, are of higher order.
An example is from the second term on the right of eq.
(\ref{K3epsilon}). It contains a term with $\langle \delta_i^3 \delta_j^2
\delta_k^2 \rangle$ in the integrand. This can be broken up into several
disconnected pieces. Most of them got canceled because of the reason
laid out in 1. above. One disconnected piece that does not
get canceled is $\langle \delta_i^3
\delta_j \rangle \langle \delta_j \delta_k^2 \rangle$. However,
applying the hierarchical relation as before, we can see
that this piece, when integrated over, is second order in
$\xibar_2^L$. Another disconnected piece is: $\langle \delta_i^3
\rangle \langle \delta_j \delta_k \rangle^2$. This deserves special
attention. It gives rise to a term in the estimation-bias of
the order of $V_T^{-2} \int dV_i dV_j [\xi_2 (i,j)]^2$, which is not
guaranteed to be much smaller than
$\xibar_2^L$ in general. However, we have checked numerically that for 
realistic power spectra, and $L$ larger than
about $10 \, {\rm h^{-1} Mpc}$, they are indeed small compared to
$\xibar_2^L$. To summarize, we can say that any products of the
two-point function where the arguments are non-degenerate (i.e. $\xi_2
[i,j]$ where $i \ne j$) can be
ignored.

3. Combining 1. and 2., it can be seen that all $\epsilon^3$ terms (or
higher) can be ignored from eq. (\ref{K3epsilon}).
The same argument applies for arbitrary $N$, and so justifies 
the dropping of $\epsilon^3$ terms from eq. (\ref{biasSN2}) or
(\ref{ratiobias}).  

4. Note that eq. (\ref{bS3epsilon}) can be derived by ignoring
$\Delta_{\xibar_2}$ and $\Delta_{\xibar_3}$ altogether. Since, the lowest order
terms in eq.
(\ref{K3epsilon}) are already of 
order $\xibar_2^L$, including $\Delta_{\xibar_2}$ and
$\Delta_{\xibar_3}$ can only give higher order
terms.

Lastly, combining eq. (\ref{bK22}), (\ref{bK3}) and
(\ref{bS3epsilon}) and substituting into eq. (\ref{biasSN2}) and
(\ref{ratiobias}), we 
obtain the net bias of the estimator $\hat S_3$:
\begin{eqnarray}
\label{biasS3noshot}
{\Delta_{S_3} \over S_3} =&&  2 {\xibar_2^L \over \xibar_2} + {4\over \xibar_2 V_T^2} \int
dV_i dV_j \xi_3 (i,j,j) \\ \nonumber
&& - {9\over \xibar_3 V_T^2} \int dV_i dV_j \xi_3 (i,i,j) - {3\over \xibar_3
V_T^2} \int dV_i dV_j \xi_4 (i,i,i,j) \\ \nonumber
&& - {2\over \xibar_2 \xibar_3 V_T^2} \int dV_i dV_j \xi_5 (i,i,i,j,j)
+ {3\over 
\xibar_2^2 V_T^2} \int dV_j dV_k \xi_4 (j,j,k,k)
\end{eqnarray}

\subsection{An Analytical Approximation}
\label{everything}

The expressions in eq. (\ref{bK22}), 
(\ref{bK3}) \& (\ref{biasS3noshot})
give the exact fractional bias in $\hat{\xibar}_2$, $\hat{\xibar}_3$
and $S_3$ to first
order in $\xibar_2^L$, excluding shot-noise. No assumption about the
two-point function $\xi_2$ itself being small has been made.
The hierarchical relation $\xi_N \sim \xi_2^{N-1}$ has only been used
for book-keeping.
We have not assumed anything about the configuration
or scale dependence of the hierarchical amplitudes.

In the present form, these expressions are not very useful as the N-point
functions, up 
to $N = 5$, are required to compute the estimation-biases.
We will approximate them as products of the two-point function
(it is from this point on, that we use the hierarchical relation
for more than simply book-keeping) using the following relation
(see \citenp{bern94b}): 
\begin{equation}
\langle \delta_i^m \delta_j^{m'} \rangle_c = c_{m m'} \xibar_2^{\, m+m'-2}
\xi_2 (i,j) \, ,
\label{cmm}
\end{equation}
where $\langle \delta_i^m \delta_j^{m'} \rangle_c$ is the connected
cosmic $m+m'$-point function (no Poisson terms), with only at most
two differing indices.
 
Putting eq. (\ref{cmm}) into eq. (\ref{bK22}), (\ref{bK3}) and
(\ref{biasS3noshot}) respectively, we obtain:
\begin{equation}
{\Delta_{\xibar_2} \over \xibar_2} = - {\xibar_2^L
\over \xibar_2} + [3 - 2 c_{12}] \xibar_2^L \, , 
\label{bK2PT}
\end{equation}

\begin{equation}
{\Delta_{\xibar_3} \over \xibar_3} = - 3 {c_{12} \over S_3} {\xibar_2^L
\over \xibar_2} + \left[ - 3 {c_{13} \over S_3} + 6 \right]
\xibar_2^L \, ,
\label{bK3PT}
\end{equation}

\begin{equation}
{\Delta_{S_3} \over S_3} = \left[2 - 9{c_{12} \over S_3} \right] {\xibar_2^L
\over \xibar_2} + \left[4c_{12} - 3 {c_{13} \over S_3} - 2 {c_{23} \over S_3}
+ 3 c_{22} \right] \xibar_2^L \, .
\label{biasS3noshotPT}
\end{equation}
It is also instructive to distinguish between the two different
contributions to  $\Delta_{S_3} / S_3$, as in eq. (\ref{intconstbias}) \&
(\ref{ratiobias}), one from the integral-constraint-biases of
$\hat{\xibar}_2$ and $\hat{\xibar}_3$ themselves:
\begin{equation}
{\Delta_{S_3}^{\rm int. cr.} \over S_3} = \left[2 - 3
{c_{12} \over S_3}\right] 
{\xibar_2^L \over \xibar_2} + \left[4 c_{12} - 3 {c_{13} \over S_3}\right]
\xibar_2^L
\label{bS3intcrF}
\end{equation}
and the other from the ratio-bias due to the division of $\hat{\xibar}_3$
by $\hat{\xibar}_2^2$:
\begin{equation}
{\Delta_{S_3}^{\rm ratio} \over S_3} =
-6 {c_{12} \over S_3}{\xibar_2^L \over 
\xibar_2} + \left[3 c_{22} - 2 {c_{23} \over S_3}\right] \xibar_2^L 
\label{bS3ratioF}
\end{equation}

In essence then, there are basically two kinds of terms in the
estimation-biases, one that does not change with the smoothing scale
$R$ (the $\xibar_2^L$ term), and the  
other that increases in magnitude as $R$ approaches the size of the
survey $L$ (the $\xibar_2^L / \xibar_2$ term). We can write this in general as:

\begin{equation}
{\Delta_E \over E} = \alpha_1 \, {\xibar_2^L \over \xibar_2} + \alpha_2 \,
\xibar_2^L
\label{E1}
\end{equation}
where $\Delta_E / E$
denotes
the fractional estimation-bias for the estimator $\hat E$, and
$\alpha_1$ and $\alpha_2$ are coefficients that depend on various 
hierarchical amplitudes, such as $S_3$ and $c_{m m'}$. 

The relation in eq. (\ref{cmm})
 is motivated by perturbation theory, and
so, strictly speaking, only holds in the weakly nonlinear regime.
But we have reasons to believe 
(from N-body work in preparation) that
the same form should work on non-linear scales, albeit
 with the coefficients $c_{mm'}$
slightly altered from the perturbative (tree order) values
(as it is known to happen with $S_N$). In the same vein,
we will use the tree order value for $S_3$ in the above estimates of
the fractional biases. This is admittedly crude for small scales, but
we will see this 
is not a bad approximation in the next section. 
For $\xibar_2$ and $\xibar_2^L$ on the other hand, we have used both the 
linear and non-linear values and found little difference
in the predicted biases. We use the linear values
in all the figures of this paper.

The perturbative values for the various hierarchical amplitudes are (\citenp{bern94b}; ignoring galaxy-bias): 
\begin{eqnarray}
S_3 &=& 34/7 + \gamma \nonumber \\
c_{12} &=& 68/21 + \gamma/3 \nonumber \\
c_{13} &=& 11710/441 + 61\gamma/7
+ 2\gamma^2/3 \nonumber \\
c_{22} &=& c_{12}^2 \nonumber \\
c_{23} &=& c_{12} c_{13} 
\end{eqnarray}
where $\gamma=\gamma(R)$:
\begin{equation}
\gamma \equiv {{d\log\xibar_2}\over{d\log{R}}}
\end{equation}
is the logarithmic slope
of  the variance. For a power-law power spectrum with spectral 
index $n$ we have: $\gamma = - (n+3)$.

Substituting the above into eq. (\ref{bK2PT}) to (\ref{bS3ratioF}),
it can be seen that, for $n$ of interests, a)
the overall 
estimation-biases in $\hat \xi_2$, $\hat \xi_3$ and $\hat S_3$ 
are negative, i.e. $\alpha_i <0$ in eq. (\ref{E1});
b) the integral-constraint-bias and ratio-bias
contributions to $\Delta_{S_3}/ S_3$ are comparable on 
small smoothing scales and c) the ratio-bias contribution to $\Delta_{S_3}/
S_3$ dominates on large scales. 
All these are illustrated in Fig. \ref{a1a2} which shows the coefficients
$\alpha_1$ (continuous line) and $\alpha_2$ (dashed line) as a function of
$\gamma$, for different estimators $\hat E=\hat{\xibar_2}, \hat S_3$.

Note that for a Gaussian model where $c_{ij}= S_J=0$,
the estimation-biases are quite different.
The coefficient $\alpha_2$ becomes $3$ for $\hat E = \hat{\xibar_2}$ so
that the bias is positive on small scales. 
Also, $\alpha_1 = 0$ and $\alpha_2 = 6$ for $\hat E = \hat{\xibar_3}$
while $\alpha_1 = 2$ and $\alpha_2 = 0$ for $\hat E = \hat S_3$,
which means the bias is always positive. Hence the Gaussian prediction
for the estimation-bias can be quite misleading (even for models with
Gaussian
initial conditions).

Finally, a word on what value to assume for $\xibar_2^L$.
As we have noted before in \S \ref{K2}, care should be taken
in dealing with the edge-effects. The expression for $\xibar_2^L$ is
given in eq. (\ref{K2L}) and (\ref{xi2R}). In obtaining 
$\hat{\xibar}_2$ and $\hat{\xibar}_3$, if one insists on using only
cells that do not overlap with the boundary, then one has to restrict
$i$ and $j$ in eq. (\ref{K2L}) over an inner region of the survey
where any cell centered within it does not cut the edges, and one
should equate $V_T$ with the volume of this inner region. On the other hand,
the $k$ and $l$ indices of eq. (\ref{xi2R}) should still range over
the whole survey volume. We will not try to compute $\xibar_2^L$
exactly.
Instead, we make use of the following observation:
adhering to the convention for $i$, $j$, $k$ and $l$ above, 
we can rewrite $\xibar_2^L$ as $\int dV_k dV_l f_k f_l \xi_2^{\rm
usmth.} (k,l)$, where $f_k \equiv V_T^{-1} \int dV_i W(k,i)$. 
The quantity $f_k$ varies between $1/V_T$, for $k$ sufficiently far
away from the edges, to $0$, for $k$ sitting on the boundary. 
If one very crudely replaces $f_k$ by its volume-average, which
is equal to $1/V_T'$ where $V_T'$ is the total volume of the survey
(everything within the boundary), it can be seen that $\xibar_2^L$ is
then simply equal to $\xibar_2$, except with a funny top-hat that
covers the whole volume of the survey.
We will further approximate this by estimating $\xibar_2^L$
using $\xibar_2$ with a spherical top-hat of size $R_L$ such that
its volume is the same as that of the survey (e.g. if the survey is
a cubical box of side-length $L$, then $L^3 = 4 \pi R_L^3 / 3$).
This is admittedly crude, but seems to be sufficiently accurate
for the N-body experiments we study, at least for a smoothing scale $R$
which is not too large compared to the size of the survey.
In practice, one might want to go back to 
the original definition of $\xibar_2^L$ (eq. [\ref{K2L}]), and
compute $\xibar_2^L$ more carefully. 

\section{Comparison with N-body Simulations}
\label{numerical}

To test our analytical predictions in the last section, 
we use simulations of two
different spatially flat
 cold dark matter (CDM)
dominated models.
One set of simulations is of the SCDM model, with $\Omega_{0}=1, h=0.5$,
  and another
is of the LCDM model with $\Omega_{0}=0.2, h=1$ and $\Omega_{\Lambda} = 0.8$.
The power spectra, P(k) for these models are
 taken from Bond \& Efstathiou \cite*{bond84}
 and Efstathiou, Bond and White \cite*{ebw92}. The shape of P(k) is
parametrized by the quantity $\Gamma=\Omega h$, so that we
have $\Gamma=0.5$ CDM and  $\Gamma=0.2$ CDM.
 Each simulation
contains $10^{6}$ particles in a box of comoving side-length $300 \Mpc$
and was run using a P$^{3}$M $N$-body code 
(\citenp{hockney81,edwf85}). 
All outputs are normalized to
$\sigma_8 =1$.
The simulations are described in more detail in Dalton et al. \cite*{dalton94}
and Baugh \etal \cite*{bge95}.
In this paper we use 10 realizations of each model for computing
ensemble averages, with error bars being
estimated from their standard deviation. 

>From each realization we extract one subsample 
within a cubical box of size $L=300 \Mpc / M$, where $M$
is taken to be an integer, $M=1,2,3,...,7$. So we have
a set of 10 realizations of subsamples for each box-size,
from $L \simeq 40 \Mpc$ to $300 \Mpc$.
We estimate moments of counts-in-cells (as in \citenp{bge95})
in each set of subsamples to study how the estimation biases of the
variance and skewness vary with survey volume.

Note the importance of using subsamples of large simulations,
rather than running simulations with an intrinsically small box-size.
The latter introduces dynamical effects due to
the missing of the large scale power, which we are not
interested in for the purpose of this paper.
Note also the importance of taking one subsample from each
realization, rather than extracting multiple subsamples from a single
realization, to ensure the statistical independence of the subsamples.
 
\subsection{The Variance $\xibar_2$}
\label{varianceN}

The results for the variance of the LCDM model are shown in Figure
\ref{x2c60all}. Because the LCDM has more power on large
scales this model shows a more pronounced integral-constraint-bias
compared to the SCDM model.
Open circles show the measured variance averaged over the 10
realizations of the 
full box ($L=300 \Mpc$). 
Filled triangles show the mean measured variance 
for the smaller boxes with the box-size
$L=300/2, 300/4, 300/5, 300/7 \Mpc$ 
as indicated in each panel. In all cases, the error-bars
represent $1-\sigma$ deviations in the measured variance
over the relevant 10 realizations.
The solid line gives the linear
perturbation theory prediction for $\xibar_2$.
(This is not perturbation theory in the sense of \S \ref{outline}, but
perturbation theory in the usual sense: an expansion in $\xibar_2$ or
the density fluctuation amplitude; to
avoid confusion, we will refer to it simply as PT.)
The agreement of the solid line and the open circles on large scales
indicate that the measured variance from the full box does not
suffered from a significant bias, for the smoothing scales shown.
The dashed line is the integral-constraint-bias prediction
in eq.[\ref{bK2PT}],
which is in excellent agreement with the simulation results.

\subsection{The Skewness $S_3$}
\label{S3N}

The results for the skewness
 are shown in Fig.
\ref{s3c150all} \& \ref{s3c60all}
for the SCDM and LCDM models. Again, because the LCDM has more power on large
scales, this model shows a larger estimation-bias.
Open circles show the mean $\langle \hat S_3 \rangle = \langle \hat{\xibar}_3
/ \hat{\xibar}_2^2 \rangle$ in 10 realizations of the
full box ($L=300 \Mpc$). 
Filled triangles show the mean $\langle \hat S_3 \rangle$ over
10 realizations for each smaller box-size:
$L=300/2$, $300/4$, $300/5$, $300/7 \Mpc$
as indicated in each panel. The squares, on the other hand,
show the mean $\langle \hat{\xibar}_3 \rangle$ divided by the mean
$\langle \hat{\xibar}_2^2 \rangle$, which is our way of isolating the
integral-constraint-bias of $\hat S_3$ (eq. [\ref{bS3intcrF}]).

The solid line corresponds to the tree-level PT prediction for
$S_3$. Its agreement with the open circles on large scales indicates
that the measured skewness from the full box does not suffer from any
appreciable estimation-bias, for the smoothing scales shown.
The short-dashed line is our analytical prediction for the
integral-constraint-bias of $\hat S_3$ (eq. [\ref{bS3intcrF}] i.e. no
ratio-bias), and is in good agreement with the simulation results (squares).
The long-dashed line is the net estimation-bias of $\hat S_3$ (eq.
[\ref{biasS3noshotPT}]), which
includes both the ratio-bias (eq. [\ref{bS3ratioF}]) and the
integral-constraint-bias (eq. [\ref{bS3intcrF}]), and should   
therefore be compared with the triangles from the simulations.
It can be seen that the ratio-bias of $\hat S_3$ (eq.
[\ref{bS3ratioF}]) always dominates on large scales, and that
the integral-constraint-bias of $\hat S_3$ is negligible except
for the smallest subsamples.

The agreement between our analytical prediction and the simulation
results is good as long as the estimation-bias is not too large.
Our analytical prediction breaks down when the bias
becomes too large, as in the case of $L=300/7 \Mpc$.
This is hardly surprising as the calculation
was done by explicitly assuming that the estimation-bias is small.
We have found a phenomenological fit to correct for this:
instead of having $\langle \hat S_3 \rangle = S_3 (1 +
\Delta_{S_3}/S_3)$,
we use
\begin{equation}
\langle \hat S_3 \rangle = S_3 \exp(\Delta_{S_3}/S_3)
\label{ansatz}
\end{equation}
where $\Delta_{S_3}/S_3$ is our linear estimate (in $\xibar_2^L$) for the
fractional bias as before 
(eq. [\ref{biasS3noshotPT}]). The higher order terms from the
exponential helps partially cancel the over-prediction due to the
linear term alone.
This ansatz is shown as a dot-dashed line in the
Fig. \ref{s3c150all} \& \ref{s3c60all}. As can be seen, there is a
reasonable agreement with the 
simulation results of both models, indicating that this is an acceptable
extrapolation. An alternative would be to go beyond linear order
in $\xibar_2^L$, and compute the estimation-biases to second order.
We will not attempt to do so here.

It should be emphasized that we have used the perturbation theory values
for quantities such as $c_{m m'}$, $S_3$ and $\xibar_2$ in the 
analytical predictions for the various estimation-biases
(eq. [\ref{bK2PT}], [\ref{biasS3noshotPT}] \& [\ref{bS3intcrF}]).
The good agreement on small scales between our analytical predictions and the
numerical results above should be seen as somewhat fortuitous.
Lacking actual measurements from simulations of the quantities $c_{m m'}$ on 
nonlinear scales, we will not attempt to do any better here.
In practice, one might want to use improved determinations of
$c_{m m'}$, etc at nonlinear scales, from simulations for example, 
in eq. (\ref{bK2PT}) to (\ref{biasS3noshotPT}), or 
even go back to their original formulations in eq. (\ref{bK22}), 
(\ref{bK3}) and (\ref{bS3epsilon}). 
But there are a few reasons why our PT-based analytical predictions
should work fairly well:
1. the terms involving $\xibar^L_2 / \xibar_2$ are only important
on large scales, and so using the PT values 
is adequate for these terms; 2. the true $S_3$ and $c_{m m'}$ change 
only slowly with the smoothing scale (i.e. the tree-order PT predictions
are not too far off); 3. most of the terms involve
ratios of $c_{m m'}$ and $S_3$, which are perhaps even slower functions
of the smoothing scale.

\section{Simulated Galaxy Catalogues and a First Step Towards Reconciliation}
\label{catalogue}

To make contact with existing measurements from actual galaxy
catalogues, we add three elements of realism to our simulations:
1. use box-sizes similar to those of surveys where some of the
measurements of the  
variance and skewness have been made; 2. introduce redshift distortions
for redshift catalogues and projection for angular-catalogues; 
3. allow for a realistic level of (sparse) sampling.
Galaxy-biasing is not implemented here, however.

We simulate CfA/SSRS volume-limited catalogues similar to those studied
in Gazta\~{n}aga \cite*{g92}, based on the LCDM model.
Redshift distortions are modeled in the usual way, with the distant
observer approximation. We consider two sets of subsamples,
taken from each of the 10 full-box ($L
= 300 \Mpc$) realizations of the LCDM model as in Fig. \ref{s3c60all}.
The first set has the same volume as the CfA/SSRS50 catalogues
in Table 1 (a cubic box of $L = 40 \Mpc$ on a side,
with a volume equivalent to that of a sphere of radius $R_L = 25 \Mpc$;
the effective depth is ${\calD} \simeq 50 \Mpc$)
and the second set has a volume similar to the CfA92 catalogue
in Table 2 (a cubic box of 
$L = 78 \Mpc$ on a side, with a volume equivalent to that of a sphere of
radius $R_L = 48 \Mpc$; the effective depth is ${\calD} \simeq 92
\Mpc$).
We have checked that the results in this section are 
essentially unchanged if, instead of a cubical box, one considers a
conical geometry which resembles that of the actual surveys.
This is in part because of the actual solid angle substended by these surveys
($\sim 1.6$; see caption of Table 1).

Following Gazta\~{n}aga \cite*{g92}, we 
concentrate on 
counts-in-cells for spherical cells 
of radius $R$ in a range between $4-22 \Mpc$, 
depending on the subsample.
The lower limit is chosen to avoid too much
shot-noise, whereas the upper limit is picked to avoid large edge-effects.  
We vary the number of particles/galaxies in each catalogue
to assess the effect of shot-noise on the estimation-biases.

Simulated angular catalogues that resemble the APM are taken from
Gazta\~{n}aga \& Bernardeau \cite*{gb98}. The power spectrum is that
measured from the APM. 
We consider two different box-sizes $L = 378 \Mpc$ and $L = 600 \Mpc$.


\subsection{The Variance $\xibar_2$ in Simulated CfA/SSRS Catalogues}

Figure \ref{x2cfa} shows the integral-constraint-bias for
$\hat{\xibar}_2$ from the simulated CfA/SSRS catalogues.
The point-symbols correspond to the values
measured from the simulated catalogues, while the lines correspond
to either values measured from the full-box LCDM simulations (long- and
short-dashed lines) or analytical predictions for
the integral-constraint-bias (solid lines).
The variance measured in the nearby sample (CfA/SSRS50) is
significantly smaller than that in the deeper sample (CfA/SSRS90),
at the smoothing scale $R = 9 \Mpc$.
(Note that for clarity, we do not show the measurements from
the deeper sample at smaller scales, but they follow those from
the full-box LCDM simulation rather
closely.)
This is due to the systematic 
bias introduced by the integral constraint, which generally gets worse
for a smaller survey volume (eq. [\ref{bK2PT}]; as shown
also in Figure \ref{x2c60all}). 
This results in the phenomenon that the amplitude of the measured
variance seems 
to increase with the sample depth, as found in several studies
(e.g. \citenp{davis88,g92,bouchet93}). 
Bouchet et al. \cite*{bouchet93} correctly attributed (a significant
part of\footnote{Luminosity segregation also plays a role here.}) this
observed phenomenon to   
a finite-volume-effect along the lines of Colombi et al.
\cite*{cbs94}. Seeing this as none other than the
integral constraint allows us to predict the size of this bias
analytically. 


Note how, on large scales ($R > 9 \Mpc$), the variance in redshift
space (long-dashed line)  
is larger than that in real space (short-dashed line), as predicted by
Kaiser \cite*{kaiser87}, 
whereas the reverse holds on small scales, because of shell crossing
and virialization (the finger-of-God).
The analytical predictions (eq. [\ref{bK22}]) for
integral-constraint-bias of both catalogues are shown as 
solid lines.
We show the predictions in real-space only, but
it can be seen that the {\it biased}-estimates of
the variance in real- and redshift-space are
in fact quite similar -- we will see this even more
clearly for the skewness.
As mentioned before, these predictions
are only approximate for non-linear scales ($R \, \approxlt \, 8
\Mpc$), because 
we have not modeled properly the non-linear values of
$c_{m m'}$ and $S_J$. Nevertheless there is an overall agreement
between the simulation results and the predictions.

Lastly, we examine the effect of shot-noise on the estimation-bias, by
sparsely sampling our catalogues (use 200 galaxies in each sub-sample
instead of the $\sim 10^4$ in CfA/SSRS50 or $\sim 10^5$ in
CfA/SSRS90). 
The effect
can be seen to be small (compare closed 
triangles with open triangles).

\subsection{The Skewness $S_3$ in Simulated CfA/SSRS and
APM catalogues}
\label{simCATS3}

Figure \ref{s3cfa} shows the results for $S_3$ in the 
simulated CfA/SSRS catalogues.
Note how $S_3$ from N-body simulations 
is closer to the real space PT prediction (dotted line), when measured
in redshift space
(long-dashed line) than
in real space (short-dashed line).

As before, we clearly see the variation of the estimation-bias with
survey volume. The bias is more significant for 
CfA/SSRS50 than for CfA/SSRS90.
The analytical prediction here comes from the phenomenological 
ansatz we introduce in \S \ref{S3N} (eq. [\ref{ansatz}]).
Only the analytical prediction for real-space is shown
(solid lines). Note how the {\it biased}-measurements
of the skewness yield very similar values in real- and 
redshift-space, 
even though the true $S_3$'s 
(i.e. measured from the full box; long and short-dashed lines) 
are quite different in the two cases, especially
on small scales. In the next section, we will take advantage
of this fact and attempt to perform a preliminary
correction of some existing (biased) measurements of $S_3$ in redshift-space
(which we assume to be very close to their real-space values),
using our real-space analytical prediction for 
the estimation-bias. In other words, lacking analytical predictions
for the redshift-space $c_{m m'}$, we use the real-space PT values
for the various hierarchical coefficients in eq.
(\ref{biasS3noshotPT}) to make corrections for measurements
that are actually done in redshift-space.

\begin{figure}[htb]
\centerline{\psfig{figure=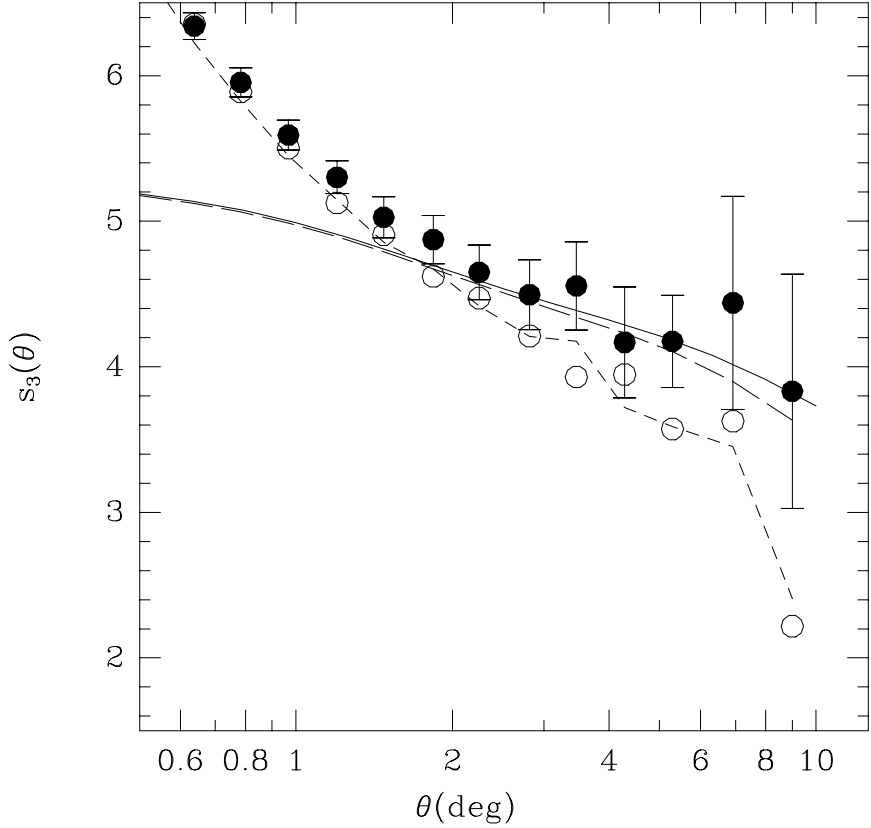,height=5.0in}}
\caption[junk]{Measurements of the projected
$S_3$ (denoted by $s_3$ here) from simulated 
APM-like angular catalogues. Closed circle: the mean of
$\hat s_3$ over 
20 angular catalogues from $L=600 \Mpc$ simulations.
Open circle: the mean of $\hat s_3$ over 20 angular catalogues
from $L=378 \Mpc$ simulations. Solid line: the tree-level perturbation
theory values of the projected $s_3$. The agreement of the solid
line with the closed circles on large scales indicate that
measurements of $s_3$ from the large box suffer from negligible
estimation-biases.
Short-dashed line: the analytical prediction for the estimation-bias
of $\hat s_3$, for the smaller simulation.
Long-dashed line: the analytical prediction for the estimation-bias
of $\hat s_3$, for the larger  simulation.}
\label{s3ang}
\end{figure}

Figure \ref{s3ang} shows the measured 2D $S_3$ in simulations
of APM-like angular  
catalogues. The results are a reproduction from
Gazta\~naga \& Bernardeau \cite*{gb98}. 
Two box-sizes are considered.
The large box with $L = 600 \Mpc$ gives measurements which 
are in good agreement with the tree-level PT predictions on large
scales, indicating that the estimation-bias is negligible in this
case, at least for $\theta \, \approxlt \, 10$ deg., e.g.
$R \, \approxlt \, 70 \Mpc$.
Note that the actual APM survey has a size even bigger than this large
box. The smaller box with $L=378 \Mpc$ results in a more appreciable
estimation-bias,. Both of this results agree with our analytical prediction
short-dashed line (long-dashed) line for the smaller (larger) box. 
The predictions can be obtained from eq.\ref{biasS3noshot} by just
replacing the  
$3D$ hierarchical amplitudes
by the $2D$ ones, as our derivation was totally general in this respect.
We use the LCDM model PT predictions with
the APM selection function and assume that $c_{ij}^{2D} \simeq 
r_{i+j} c_{ij}^{3D}$, with where $r_{i+j} \simeq 1$ as given in  
Gazta\~{n}aga \cite*{g94}. These results are not very sensitive to
the exact values of $r_{i+j}$.

\subsection{A Preliminary Reconsideration of Some
Existing Measurements from Actual Galaxy
Surveys} 
\label{firststep}

\subsubsection{The Variance $\xibar_2$ in the Actual IRAS, CfA and
SSRS Surveys} 
\label{firststepXI2}

Any corrections of existing measurements of $\xi_2$ based on 
eq.(\ref{bK2PT}) are necessarily model-dependent, because
values for quantities such as $c_{12}$, $\xibar_2^L$ and
even $\xibar_2$ 
itself need to be assumed. All of them vary with the amount and nature
of galaxy-biasing.
Instead of conducting a detailed analysis covering many possible models,
we ask the following simpler question: 
assuming that the LCDM or SCDM model
 for the shape of $\xibar_2$ are the true ones, 
 and assuming the PT-theory (real-space) values for
$c_{12}$, what would be the corrected
measurements of $\xibar_2$ for the IRAS,
CfA and SSRS catalogues, using eq.(\ref{bK2PT}),
assuming no galaxy-biasing? 


Fig. \ref{varD} shows 
the correlation length $R_0$,
defined as $\hat{\xibar_2}(R_0)=1$, 
as a function of $R_L$,
the equivalent radius of the corresponding subsample.
Open circles (squares) correspond to the values 
of the CfA (SSRS) volume limited subsamples at the bottom of Table 1.
Filled squares correspond to the values in the IRAS 1.2 Jy
volume limited subsamples
by Bouchet etal (1993). The lines show the predictions
for the measured $R_0$ taking into account the
integral-constraint-bias for $\hat{\xibar_2}$.
We adopt the shape of the linear LCDM (continuous line)
or the linear SCDM (dashed line) power spectrum to estimate
all quantities, $\xibar_2^L$, $\xibar_2(R)$ and $c_{12}$, in eq.(\ref{bK2PT}).
This is only approximate as we do not take into account redshift
distortions or non-linearities in the predictions, but note
that in Fig. \ref{x2cfa} we have found this approximation
to be good.

The amplitude of the linear power spectrum is 
chosen to give the best fit to the data points in Fig. \ref{varD}.
The best fit values for the LCDM model are
 $R_0 \simeq 10.0 \Mpc$ (where $R_0$ here is the $R_0$ we infer
from the best-fit amplitude when matching our
integral-constraint-prediction with the data points)
for the CfA/SSRS (top continuous line) which has a joint $\chi^2=5.0/6$,
 and  $R_0 \simeq 6.5 \Mpc$
for IRAS (bottom continuous line) which has a $\chi^2=37/20$.
 The best fit values for the
SCDM model are
 $R_0 \simeq 8.0 \Mpc$ for the CfA/SSRS (top dashed line)
 which has  a $\chi^2=11.4/6$,
 and  $R_0 \simeq 6.1 \Mpc$
for IRAS (bottom dashed line) which has a $\chi^2=27/20$.
Note that the SCDM model gives a poorer fit to the CFA/SSRS data,
while IRAS is compatible with both models.
These values are to be compared with
the  mean values (which are usually taken to give the  {\it true} amplitude):
$R_0 \simeq 8 \Mpc$ for the CfA/SSRS and  $R_0 \simeq 5.5 \Mpc$
for IRAS. Note that our correction for $R_0$ here ignores
galaxy-biasing. Under such an assumption, the large value of $R_0
\simeq 10 \Mpc$ (ie 
$\sigma_8 \simeq 1.2-1.3$) found in the CfA/SSRS
seems difficult to reconcile with the
amplitude infered from the angular APM Galaxy catalogue: $\sigma_8 < 1.08$
(see \citenp{g95}). To be strictly consistent, we should go
back and allow for the effect of biasing on our correction for $R_0$:
we would not attempt to do so here.
Note also that the larger the volume limited subsample the brighter the 
absolute magnitudes of galaxies it contains; our 
correction for $R_0$ implicitly ignores luminosity segregation
i.e. that the intrinsic 
clustering does not change significantly as a function of the absolute
magnitude of the galaxies. A direct
measurement in redshift space in the Stromlo-APM
Catalogue gives $\sigma_8=1.1 \pm 0.1$ for the
brightest sample with $M_{b_j}<-20$
(sample d. in Table 3 in Loveday \etal \cite*{lov96}). This
Stromlo-APM subsample contains galaxies with similar absolute
magnitudes 
to those of the CfA92 sample, where $M_{b_0}<-20.3$, 
given that $b_0 \simeq b_j -0.3$ \cite{dg98}.
Thus, our analysis indicates a small relative
galaxy-bias between the CfA/SSRS and the APM galaxies that seems 
not attributable entirely to luminosity segregation.

\subsubsection{The Skewness in the Actual CfA, SSRS and APM Catalogues }

Again here, 
it is clear that any corrections of existing measurements of $S_3$ based on 
eq. (\ref{biasS3noshotPT}) are necessarily model-dependent, because
values for quantities such as $c_{m m'}$, $\xibar_2$, $\xibar_2^L$ and
even $S_3$ 
itself need to be assumed. All of them vary with the amount and nature
of galaxy-biasing.
Instead of conducting a detailed analysis covering many possible models,
we ask the following simpler question: 
assuming different shapes for the power spectrum all normalized
to $\sigma_8=1$,
 and assuming the PT-theory (real-space) values for
$c_{m m'}$ and $S_3$, what would be the corrected
measurements of $S_3$ for the 
CfA and SSRS catalogues, using eq. (\ref{biasS3noshotPT}),
or more appropriately, its extension in eq. (\ref{ansatz}),
if no galaxy-bias is assumed? 

Note that
here, we are taking advantage of the finding in \S \ref{simCATS3}
(Fig. \ref{s3cfa})
that the biased-measurements of $S_3$ yield very similar values
in real- and redshift-space, and so we can simply apply the
correction formulated in {\it real}-space to the measurements in {\it
redshift}-space. On large scales, this is a safe assumption because
even the true $S_3$'s are very similar in real- and
redshift-space. On small scales, this assumption remains
to be further scrutinized. Obviously, this assumption must break
down when the survey-size is large enough (c.f. short- and long-dashed
lines in Fig. \ref{s3cfa}). 

\begin{figure}[htb]
\centerline{\psfig{figure=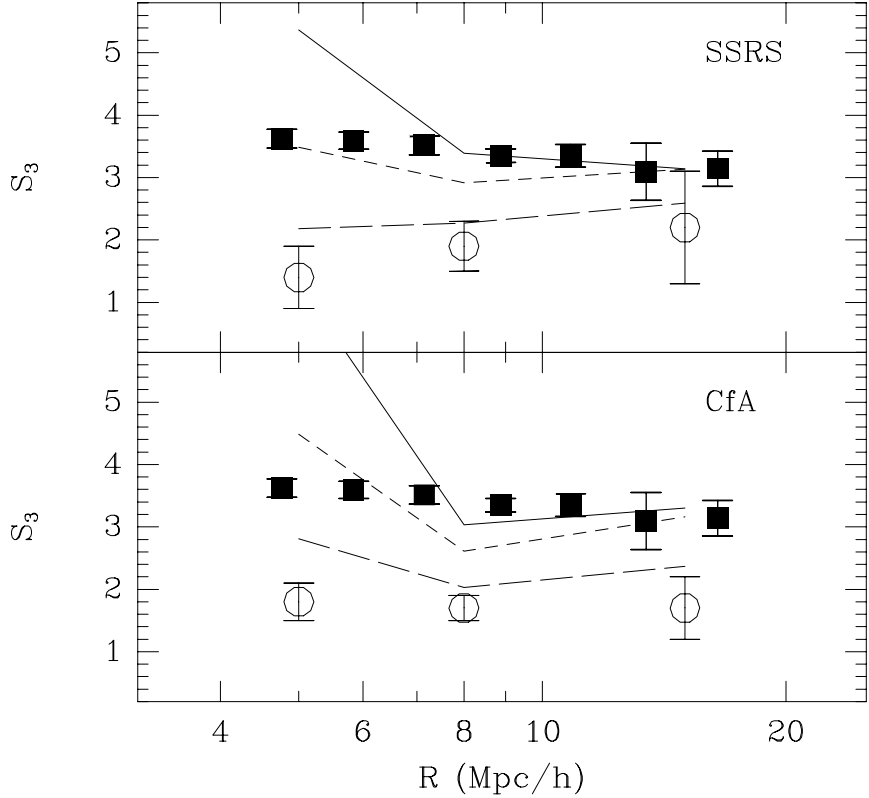,height=5.0in}}
\caption[junk]{Opened circles show the uncorrected values of $S_3$
in the SSRS
(top) and CfA (bottom) subsamples at the end of Table 1. These should
be compared to
the deprojected APM values (filled squares). The lines
show the values corrected for the estimation-bias assuming
different power-spectrum to estimate the variances
$\xibar_L$ and $\xibar_2(R)$ in eq. (\ref{biasS3noshotPT}).
The LCDM model prediction is shown as a short-dashed line,
the linear SCM model as a long-dashed line and  
the linear APM-like model as a continuous line.
All cases are normalized to $\sigma_8=1$.
The corrections are larger for smaller scales where
the measured values are from smaller sub-samples. The SCDM predictions
are much lower as it has less power on large scales.}
\label{s3bias}
\end{figure}

We will concentrate on the last six measurements in Table \ref{s3s4}. 
These published values of $S_3$ (\citenp{g92}) are the mean values
in the corresponding range of smoothing scales shown in the 4th
column. Here we will 
associate each with the mean value of $R$ in the respective range
of scales i.e. 
$R=5, 8$ \& $15 \Mpc$ with $S_3 =1.8, 1.7$ \& $1.7$ from
the CfA50, CfA80 \& CfA92 catalogues on the one hand, and
$S_3=1.4, 1.9$ \& $2.2$ from the SSRS50, SSRS80 \& SSRS115 catalogues
on the other. 
The corrected values of $S_3$, adopting the assumptions stated above,
are shown as lines in Fig. \ref{s3bias}.
The correction is larger for the smaller scales which
correspond to sub-samples of a smaller size. The SCDM predictions
are much lower as it has less power on large scales.


If there is no biasing, we can assume that the
APM-values for $\xibar_2$ is close to the  true ones, which is
is reasonable given the size of the APM (see e.g. Fig.
\ref{s3ang}). In this case we should concentrate on 
the continuous line in Fig. \ref{s3bias}.
At the larger smoothing scales, namely $8$ \& $15 \Mpc$, the corrected
values are quite consistent with the APM-values.
However, at the smallest smoothing scale of $5 \Mpc$,
the corrected $S_3$'s are significantly higher than the
APM-values. Three points should be noted here.
First, at this smoothing scale, the correction is so large that
the validity of the ansatz expressed in eq. (\ref{ansatz}) might
be called into question. Second, the implicit assumption that
the {\it biased-} measurements of $S_3$ in real- and redshift-space
yield similar values should be checked using more simulations.
Third, it is in fact well known that 
the {\it small-scale}
$S_3$ (of the mass) one would infer from an N-body simulation with an APM-like
power spectrum and Gaussian initial conditions is larger
than the measured $S_3$ from the APM survey \cite{bg96}. 
One possible interpretation is a scale-dependent galaxy-bias,
which tends to diminish on large scales but becomes significant
on small scales.


On large scales $R \, \approxgt 8 \, \Mpc$, we can
safely say that most of the discrepancies in existing measurements of
$S_3$ from the CfA/SSRS/APM catalogues can be explained by an
estimation-bias. Remaining differences are attributable 
to a) a small relative galaxy-bias between the different surveys
on large scales; b) redshift-distortions; c) deprojection effects
(see \citenp{gb98}) and d) sampling fluctuations. 
In fact, our analysis as shown in Fig. \ref{varD} does support
the existence of a galaxy-bias between the CfA/SSRS and the APM.
To be strictly consistent, we should have taken this into
account in our ``correction'' of the CfA/SSRS values for $S_3$.
Doing so is beyond the scope of the present paper.
Nonetheless, it should be emphasized that the amount of galaxy-bias
one would infer based on measurements of $S_3$ is reduced
if the estimation-bias is taken into account.
A careful assessment would require the inclusion of all 
the above effects, and ideally, a re-measurement of $S_3$ from
the different catalogues using methods that are perhaps
less prone to the ratio-bias. We hope to pursue these
in a future paper.

\section{Discussion}
\label{conclude}


The main results of this paper are summarized in eq. (\ref{bK22}), 
(\ref{bK3}) \& (\ref{biasS3noshot}), with the associated useful approximations
given in eq. (\ref{bK2PT}), (\ref{bK3PT}) \& (\ref{biasS3noshotPT}). 
Together they tell us the estimation-biases associated with the
standard estimators for the variance $\xibar_2$, the third
cumulant $\xibar_3$, and the skewness $S_3$
(eq. [\ref{biashatKN}] \& [\ref{biasSN}]). 
The calculation is based on an expansion in the small parameter
$\xibar_2^L$ (eq. [\ref{K2L}]), which is the variance smoothed on the
scale of the 
survey, but otherwise does not assume or require the smallness of
the variance on the scale of interest $R$.

>From eq. (\ref{bK2PT}), (\ref{bK3PT}) \& (\ref{biasS3noshotPT}), it 
can be seen that the standard estimators are all asymptotically
unbiased, in the sense that for a given smoothing scale $R$,
the estimation-biases tend to zero as the survey-size increases. On
the other hand, 
for a fixed survey-size, the estimation-biases become large
as $R$ approaches the size of the survey. 

\begin{figure}[htb]
\centerline{\psfig{figure=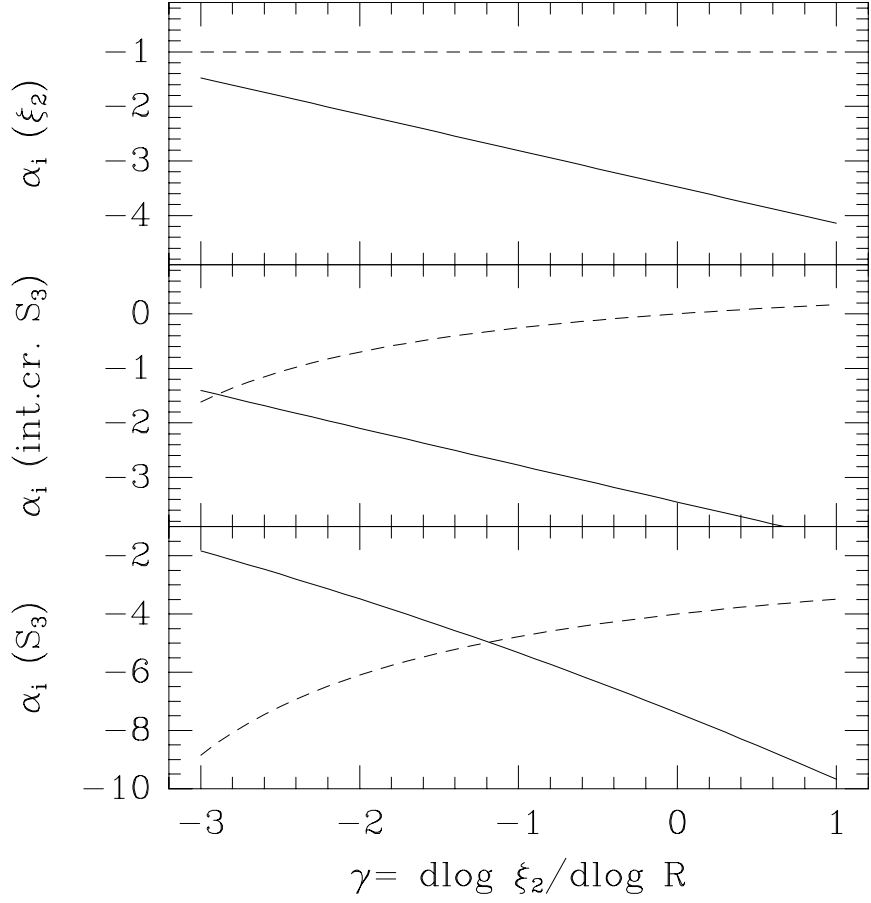,height=5.0in}}
\caption[junk]{Coefficients $\alpha_1$ (continuous line) and $\alpha_2$
(dashed line) for the fractional estimation-bias of estimator $\hat E$:
${\Delta_E \over E} = \alpha_1 {\xibar_2^L \over \xibar_2} + \alpha_2
\xibar_2^L$ (eq. [\ref{E1}]), where $\xibar_2$ is the variance
smoothed on scale $R$ and $\xibar_2^L$
is the variance smoothed on the scale of the
survey. The top panel shows the
coefficients for the variance $\hat E= \hat{\xibar_2}$
(eq.[\ref{bK2PT}]), and the 
bottom panel shows the coefficients for the skewness $\hat E= \hat
S_3$ (eq.[\ref{biasS3noshotPT}]), 
while the middle panel shows the integral-constraint-bias contribution
to $\hat E= \hat S_3$ (eq. [\ref{bS3intcrF}]).}
\label{a1a2}
\end{figure}

There are two types of terms in the fractional
estimation-biases, one dependent on the smoothing scale $R$,
being proportional to $\xibar_2^L / \xibar_2$ where $\xibar_2$ is the variance
smoothed on scale $R$, and the other not, being proportional to
$\xibar_2^L$ only. In other words, a general form for the
estimation bias of an estimator $\hat E$ can be represented by eq.
(\ref{E1}) where 
$\alpha_1$ and $\alpha_2$ are coefficients that depend on the various 
hierarchical amplitudes, such as $S_3$ and $c_{m m'}$. 
For reasonable choices of the parameters $S_3$ and $c_{m m'}$, the
estimation-biases are negative (see Fig. \ref{a1a2}). 
The magnitude of the estimation-biases can be surprisingly large,
especially for 
$E = S_3$, because of the large coefficients multiplying $\xibar_2^L$ or
$\xibar_2^L / \xibar_2$. 

\begin{figure}[htb]
\centerline{\psfig{figure=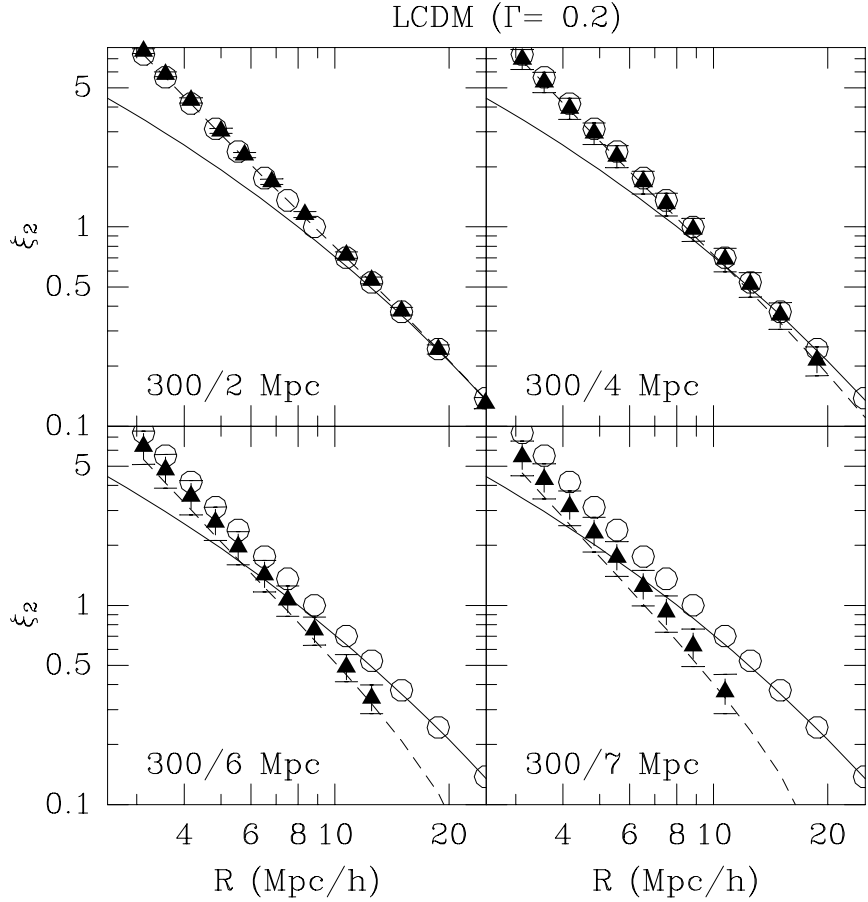,height=5.0in}}
\caption[junk]{
The integral-constraint-bias of the variance-estimator
${\hat{\xibar}}_2$ in  
the LCDM model as a function of the smoothing scale $R$.
Open circle: the average measured variance, $\langle
\hat{\xibar}_2\rangle$, computed from 10
realizations of the full box ($L=300 \Mpc$).
Filled triangle: the average measured variance computed using
10 subsamples, each extracted from each realization of the full box ($L = 300
\Mpc/M$, for $M = 3, 4, 5$ \& $7$ as labeled for each panel).
The $1-\sigma$
error-bars are computed from the 
dispersion of the measured variance around the mean over the 10 respective
realizations. 
Short-dashed line: analytical 
prediction for the integral-constraint-bias (eq. [\ref{bK2PT}]).
Solid line: tree-level PT prediction for 
$\xibar_2$.} 
\label{x2c60all}
\end{figure}

\begin{figure}[htb]
\centerline{\psfig{figure=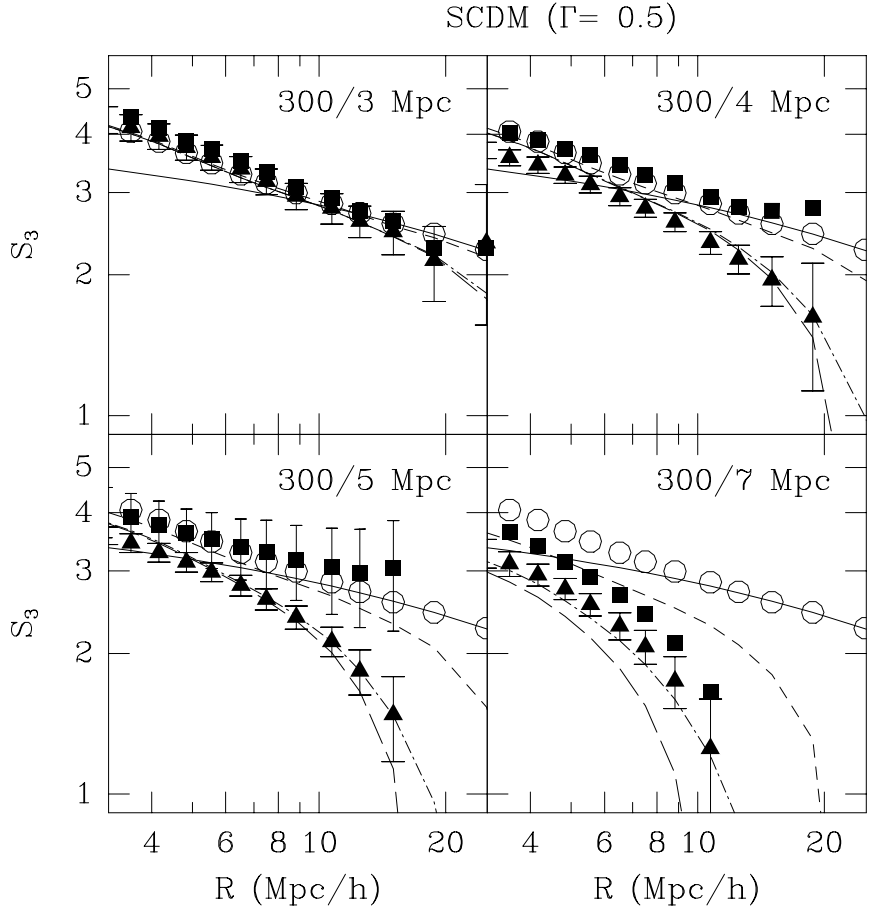,height=5.0in}}
\caption[junk]{The estimation-bias for $\hat S_3$
in the SCDM model as a function of the smoothing scale $R$. Open circle: 
the mean of $\hat S_3$ over 10 realizations of the full box
($L=300 \Mpc$). Filled triangle: the mean of $\hat S_3$
over the corresponding subsamples of each realization
of the full box (subsample box-sizes are
$L= 300 \Mpc/M$, for $M = 2, 4, 5$ \& $7$ as labeled for each
panel).
Filled square: the mean of $\hat{\xi}_3$ divided by the
mean of $\hat \xi_2^2$ over the corresponding 10 subsamples
(i.e. the integral-constraint-bias only).
The error-bars are computed from the standard deviation of the measured
values over the respective 10 realizations; for clarity, error-bars
for the squares are only shown in one panel.
Short-dashed line:
the analytical prediction for the integral-constraint-bias of $\hat
S_3$ (eq. [\ref{bS3intcrF}]). Long-dashed line: 
the analytical prediction for the net
estimation-bias of $\hat S_3$ (eq. [\ref{biasS3noshotPT}]). 
Solid line: the tree-level PT prediction for 
$S_3$. Dot-dashed line: a phenomenological ansatz for the net
estimation-bias when it becomes large (eq. [\ref{ansatz}]).}
\label{s3c150all}
\end{figure}

\begin{figure}[htb]
\centerline{\psfig{figure=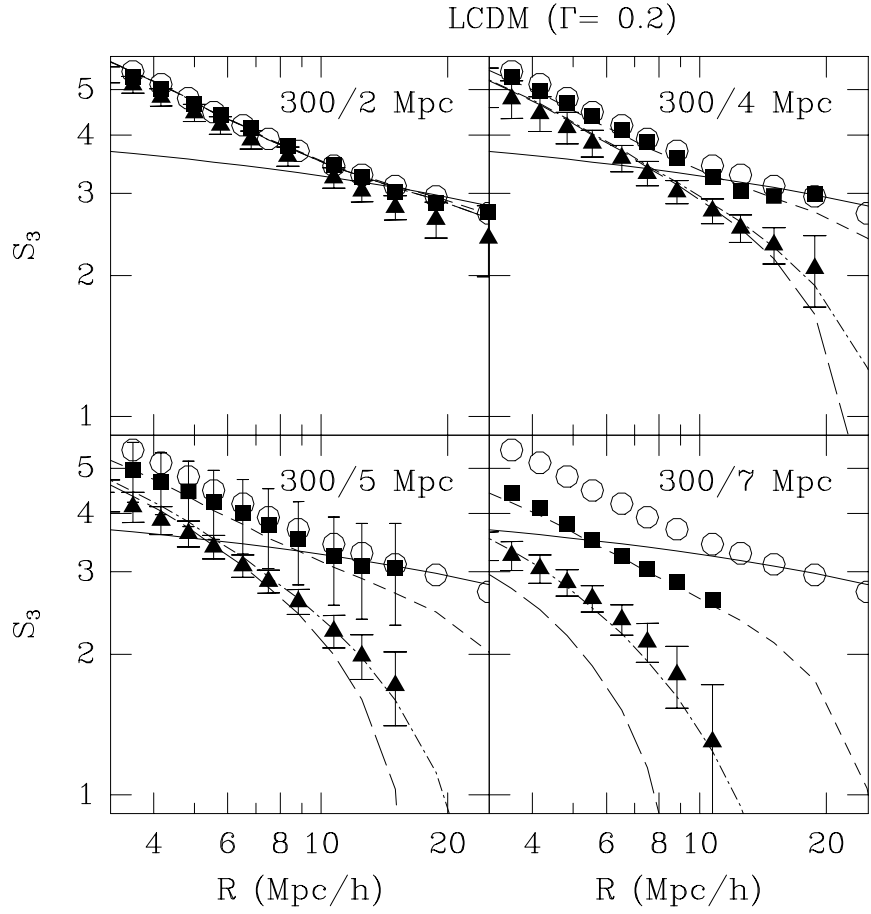,height=5.0in}}
\caption[junk]{The estimation-bias for $\hat S_3$ in the LCDM model, labeled in
exactly the same way as in Fig. \ref{s3c150all}.}
\label{s3c60all}
\end{figure}

\begin{figure}[htb]
\centerline{\psfig{figure=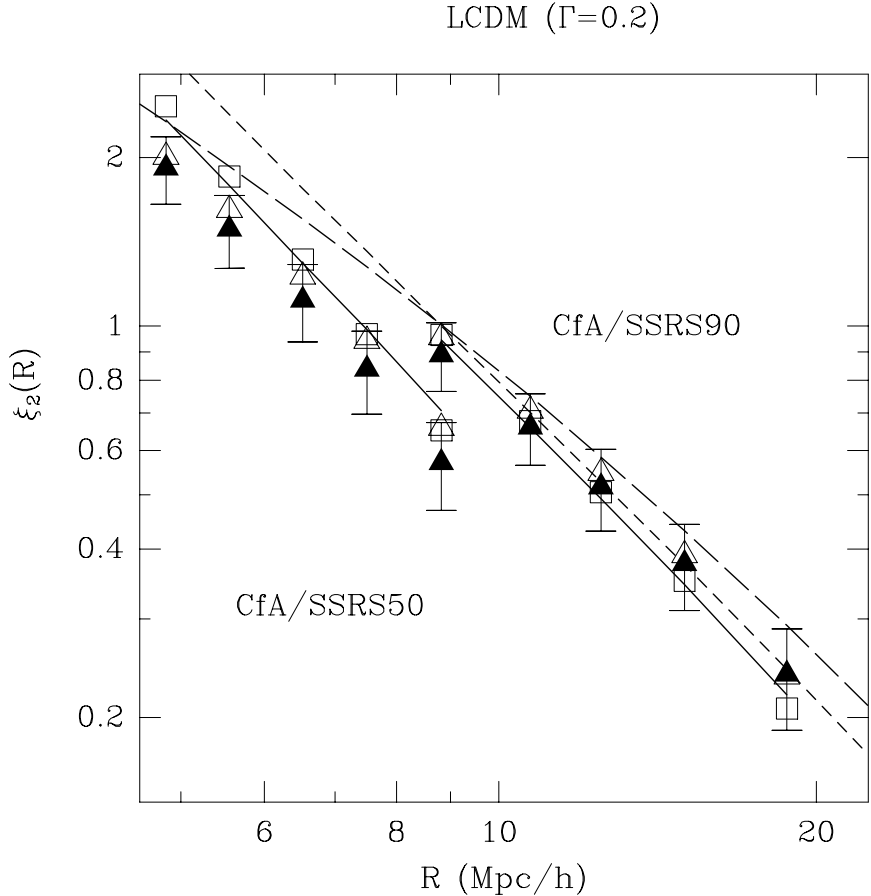,height=5.0in}}
\caption[junk]{
Measurements of $\xibar_2$ from simulated
CfA/SSRS catalogues based on earlier LCDM simulations
with $\Gamma=0.2$ (see Fig. \ref{x2c60all}).
Short- and long-dashed lines shows the LCDM simulation results in
real and redshift space respectively for the full box ($L = 300
\Mpc$). The two sets of point-symbols (one on the left, one on the right)
depict the measurements from two different volume-limited
simulated catalogues:
a) CfA/SSRS50 (for $R \le 9 \Mpc$), limited to ${\cal{D}} \sim 50 \Mpc$
b) CfA/SSRS90 (for $R \ge 9 \Mpc$), limited to ${\cal{D}} \sim 90 \Mpc$. 
Square: real space, full sampling ($\sim 10^4$ galaxies for
CfA/SSRS50 \& $\sim 10^5$ galaxies for CfA/SSRS90). Open triangle:
redshift space, full sampling. 
Closed triangle with error-bars: redshift space, sparse sampling (200 galaxies).
The solid line shows our analytical predictions
of the integral-constraint-biases in real-space (eq.
[\ref{bK2PT}]) for the respective catalogues. Note that 1. the larger
simulated catalogue yields a larger 
measured variance at $9 \Mpc$; 2. 
the sparse-sampling does not significantly affect the mean determination
of the variance; 3. the real-space and redshift-space {\it biased}-estimations
of the variance give very similar values.}
\label{x2cfa}
\end{figure}

\begin{figure}[htb]
\centerline{\psfig{figure=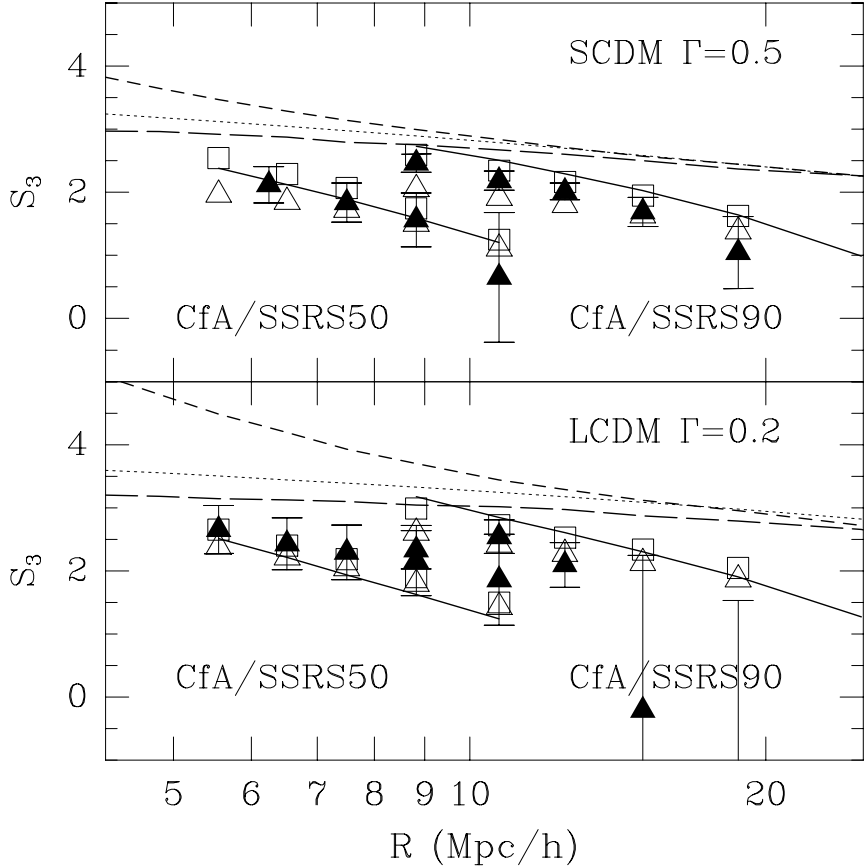,height=5.0in}}
\caption[junk]{Measurements of $S_3$ from simulated
CfA/SSRS catalogues based on earlier CDM simulations
with $\Gamma=0.5$ (top panel) and $\Gamma=0.2$ (bottom panel) (see
Fig. \ref{x2c60all}). 
Dotted line shows the tree-level PT prediction
for each model (e.g. $S_3=34/7+\gamma$).
Short- and long-dashed lines show the measured $S_3$ from simulations
of the full-box ($L = 300 \Mpc$) in
real and redshift space respectively. The point-symbols show
measurements
from  two sets 
of volume-limited simulated catalogues:
a) CfA/SSRS50 (left), limited to ${\cal{D}} \sim 50 \Mpc$
b) CfA/SSRS90 (right), limited to ${\cal{D}} \sim 90 \Mpc$. 
Square: real space, full sampling ($\sim 10^4$ galaxies for
CfA/SSRS50 \& $\sim 10^5$ galaxies for CfA/SSRS90). Open triangle:
redshift space,  full sampling.
Closed triangle: redshift space, 200 galaxies.
The solid line shows our prediction
of the estimation-bias for $\hat S_3$, using the
ansatz introduced in \S \ref{S3N}, in real
space.  Note how the {\it biased}-estimations (from the
simulated catalogues) of the skewness
give very similar values in real- and redshift-space,
even though the true $S_3$'s (i.e. measured from the
full box) are quite different in the two cases, especially at
small smoothing scales.
}
\label{s3cfa}
\end{figure}

Our analytical predictions are borne out by numerical experiments
discussed in \S \ref{numerical}. Examples can be found in
Fig. \ref{x2c60all}, \ref{s3c150all} \& \ref{s3c60all}. 
In cases where the estimation-biases are so large
that the higher order terms in our expansion in $\xibar_2^L$ become
important, we show that a simple ansatz works reasonably well (eq.
[\ref{ansatz}]). 

In the case of $S_3$, we distinguish between two types of
contributions to its estimation-bias. The standard
estimator for $S_3$ is $\hat S_3 = \hat{\xibar_3}/\hat{\xibar_2}^2$.
Part of the bias arises from the 
biases of the estimators $\hat{\xibar_3}$ and $\hat{\xibar_2}$
themselves -- this is the integral-constraint-bias (eq.
[\ref{bS3intcrF}]). The second part arises 
from the particular nonlinear combination of these two estimators,
which we dub to be the ratio-bias (eq. [\ref{bS3ratioF}]), or more
precisely, the 
nonlinear-estimation-bias. 
It turns out the second always dominates on large scales.

\begin{figure}[htb]
\centerline{\psfig{figure=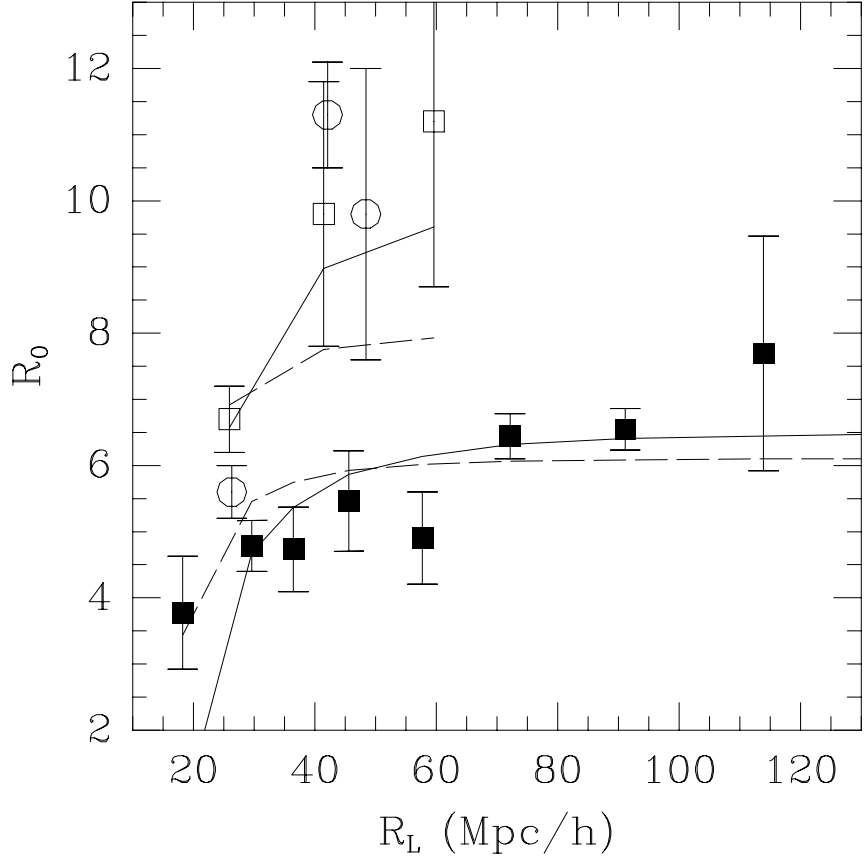,height=5.0in}}
\caption[junk]{The correlation length,
defined as $\hat{\xibar_2}(R_0)=1$, as a function of $R_L$
the equivalent radius of the volume limited subsample.
Open circles (squares) correspond to the values 
of the CfA (SSRS) subsamples at the bottom of Table 1.
Filled squares correspond to the values in the IRAS 1.2 Jy
by Bouchet \etal (1993). 
The continuous (dashed) line shows the prediction for
the measured $R_0$, which takes into account the
integral-constraint-bias, for the LCDM (SCDM) model, whose
power-spectrum-amplitude 
amplitude is adjusted to fit the data points.}
\label{varD}
\end{figure}

We present a preliminary attempt to correct some existing
measurements of the variance and skewness 
in \S \ref{catalogue}. Our main conclusions are
a) the apparent increase of the measured variance with survey depth
observed by some authors can be nicely explained by the
integral-constraint bias (Fig. \ref{varD}; see e.g.
\citenp{davis88,g92,bouchet93}); 
b) our analysis indicates a small relative
galaxy-bias between the CfA/SSRS and the APM galaxies, 
c) the APM survey should give small estimation-biases for the
standard estimator for $S_3$ on scales of interest (see Table
\ref{s3s4}); 
d) on large scales $R \, \approxgt 8 \, \Mpc$, most
of the differences between the measured skewness from the
CfA/SSRS and the APM surveys can be attributed to the estimation-bias;
however, a more careful analysis, taking into account redshift-space
distortions, relative galaxy-bias (such as due to luminosity
segregation) and deprojection effects,
is necessary to access the significance of the remaining differences,
and whether they are due to pure sampling fluctuations.

Two areas clearly warrant further investigations.
First, for the purpose of future surveys such as the SDSS or the AAT
2dF, the $\xibar_2$-independent terms in the estimation biases
(the terms associated with $\alpha_2$ in eq. [\ref{E1}]) will probably be
unimportant, but the 
$\xibar_2$-dependent terms (the $\alpha_1$-terms)
can always become important for a sufficiently large
smoothing scale $R$. 
\footnote{However, for most current models of
the power spectrum, as $R$ increases $\gamma$ becomes
more negative and the coefficient $\alpha_2$ in $S_3$
becomes smaller (see Fig. \ref{a1a2}).}
It would therefore be
good to have an idea of what that scale is, not just for the
variance and the skewness, but also for $S_N$ where $N > 3$. 
We will present results for a general N in a separate paper.

Second, we have focused in this paper exclusively on the standard estimators
for $\xibar_N$ and $S_N$, where
$\xibar_N$ is estimated by the standard counts-in-cells technique
and $S_N$ is estimated by
taking the appropriate ratio of ${\hat{\xibar}}_N$ and ${\hat{\xibar}}_2$.
There are probably other estimators that suffer from smaller
estimation-biases. For instance, Kim \& Strauss \cite*{kimstrauss98}
recently introduced an interesting method to obtain $S_3$ by fitting
the one-point probability distribution of counts (PDF) using
an Edgeworth series. They obtained values higher than previous
measurements from the same surveys (see Table \ref{s3s4}) indicating
that their method is less susceptible to an estimation-bias.
An extension of their method using a PDF which is
better behaved in the presence of more significant nonlinearities
is worth pursuing. 

Another possibility which might reduce the ratio-bias
in $\hat S_N$ due to the division of ${\hat{\xibar}}_N$ by
${\hat{\xibar}}_2^{N-1}$: instead of
dividing to obtain an estimate of $S_N$, fit a curve 
parametrized in some form to the two-dimensional plot of
${\hat{\xibar}}_N (R)$ and ${\hat{\xibar}}_2 (R)$ at each smoothing scale $R$.
This is akin to, for instance, how the
Hubble constant is usually measured: instead of dividing some estimate
of velocity by 
some estimate of distance for each data point followed by averaging, a linear
$\chi^2$ fit 
to all points is performed. This method
might give a smaller ratio-bias, but it has to be tested.
This procedure has in fact been carried out before
by e.g. Gazta\~{n}aga \cite*{g92} and Bouchet et al. \cite*{bouchet93},
who obtained values of $S_3$ close to that using
the standard estimator. But a different parametrization
of the relation between ${\hat{\xibar}}_N (R)$ and ${\hat{\xibar}}_2 (R)$
(in other words, taking into account carefully the change of $S_N$
with $R$) might yield different 
results. 

Perhaps the most important and obvious lesson of our investigation
here is that nonlinear combinations of estimators
should be used with caution.
The only sure-fire way of avoiding an estimation-bias is through
Monte-Carlo simulations, 
such as those performed in \S \ref{numerical}. 
There is nothing novel about this point, except that
its importance has not been sufficiently emphasized in
measurements of certain large scale structure statistics, 
such as $\xibar_N$ and $S_N$ discussed here.

There are also strong interests in measuring such quantities outside
galaxy surveys, e.g. for the transmission distribution of quasar
spectra and for the convergence distribution in weak-lensing maps.
The skewness of the former, for instance, provides a test of the
gravitational-instability picture of the Lyman-$\alpha$ forest
(\citenp{hui98}), 
while the skewness of the latter is a sensitive probe of cosmology
(\citenp{bernardeau97}).
Measurements in these areas require similar caution as in the
case of galaxy surveys. Our methodology in deriving the
estimation-bias for the standard estimator of skewness
(ratio of the third cumulant to the second cumulant squared)
could be adapted for such measurements (in the case of weak lensing,
we learned as this work was being completed that a related
calculation was done by \citenp{bernardeau97}). 
An important ingredient of our calculation is the use
of the hierarchical relation $\xi_N \sim \xi_2^{N-1}$ in keeping
track of the ordering, which might have to be modified for
these other applications.

Also, for these applications, as well as for less conventional
galaxy surveys such as the Lyman-break galaxy surveys at high
redshifts (\citenp{steidel98}), one often has available several
independent fields for which a simple and obvious method should help to
reduce much of the ratio-bias of the skewness: measure
the third moment and the second moment for each field, separately
average each of them over all fields, and only then does one
combine these averaged moments to estimate the skewness.

A related suspect of a similar estimation-bias is the measurement of
$Q_N$, defined as the ratio of the N-point correlation to the
sum of suitable permutations of products of the two-point functions
(\citenp{fry84}). The configuration dependence of $Q_3$, for instance,
provides an elegant test of the galaxy-bias (\citenp{fry94}).
Common ways of estimating it, where
estimators are divided by each other, are susceptible
to the ratio-bias just as in the case of $S_N$. 
Another possibly problematic statistic is the ratio of
the quadrupole to monopole power in redshift-space, or the
ratio of the monopole power in redshift-space to that in real-space,
which is often used to estimate the parameter $\beta = \Omega^{0.6}
/b$ (see review by \citenp{hamilton97}).
An examination of published estimates of $\beta$ show
a large scatter even from the same surveys,
with maximum likelihood 
methods yielding consistently higher values (\citenp{hamilton97}), suggesting
an estimation-bias of some sort might be lurking here.
It is likely, however, that such attempts to measure
$\beta$ are at least equally, if not more strongly, affected by our poor
understanding of translinear distortions. We hope to pursue some of the above
issues in future work.

We are grateful to Joshua Frieman for useful discussions.
This work was in part supported by the DOE and the NASA grant NAG 5-7092 at
Fermilab, and by a NATO Collaborative Research Grants 
Programme CRG970144 between Fermilab and IEEC. EG acknowledges support by IEEC/CSIC, 
and by DGES (MEC),  project PB96-0925.

\vspace{1.0in}

\pagebreak

\begin{table}[htb]
\centering
\begin{tabular}{|c|c|c|c|c|c|c|} 
 \hline   \rule[-0.6ex]{0mm}{3.2ex} \makebox[3em]{$R_0$ } &
\makebox[4em]{$S_3$} &\makebox[4em]{$S_4$} 
 &\makebox[3em]{Scales} &\makebox[6em]{Sample}&
\makebox[3em]{$R_L$}
& \makebox[10em]{Reference} \\* \hline
\rule[-0.6ex]{0mm}{3.2ex} Mpc/h & $(3 Q_3)$  & $(16 Q_4)$& Mpc/h&  & 
Mpc/h & 
\\* \hline \rule[-0.6ex]{0mm}{3.2ex}
7 & $3.9 \pm 0.6$  &  ---     & 0.5-8 &LICK  & 210 & Groth \& Peebles 
1977 
\\* \hline \rule[-0.6ex]{0mm}{3.2ex} 
7 & ---  & $48 \pm 7$  & 0.5-4 & LICK &  210 & Fry \& Peebles 1978   
\\* \hline \rule[-0.6ex]{0mm}{3.2ex}
7 & $4.3 \pm 0.2$  &  $31 \pm 5$  & 0.5-4  & LICK  & 210 & Szapudi et al.
1992 
\\* \hline \rule[-0.6ex]{0mm}{3.2ex}
8 & $4 - 7$  &  $30 - 150$   & 0.1-7 & EDSGC & 240 & Szapudi et al. 1996 
 \\* \hline \rule[-0.6ex]{0mm}{3.2ex}
8 & $4.1 \pm 0.1 $  &  $37 \pm 2$   & 0.3-2 & APM (17-20) & 380 & 
Gazta\~naga 1994 
 \\* \hline \rule[-0.6ex]{0mm}{3.2ex}
8 & $3.2 \pm 0.2 $  &  $33 \pm 4$   &  7-30 & APM (17-20) & 380 &  ``  
 \\* \hline \rule[-0.6ex]{0mm}{3.2ex}
8 & $3.1 \pm 0.5 $  &  $25 \pm 7$   & 0.5-50 & APM (mean) & 150-380 & 
Szapudi et.al 1995
\\* \hline \hline \rule[-0.6ex]{0mm}{3.2ex}
10 & $2.2 \pm 0.2$  &  $10 \pm 3$   & 4-20  & IRAS 1.2Jy  & 90 & Meiksin 
et al. 1992 
\\* \hline \rule[-0.6ex]{0mm}{3.2ex}
6 & $2.4 \pm 1.1$  &  $11 \pm 13$   &  2-20 & IRAS 1.9Jy & 35-60 & Fry \& 
Gazta\~naga 1994 
\\* \hline\rule[-0.6ex]{0mm}{3.2ex}
7 & $2.1 \pm 0.6$  &  $7.7 \pm 5.2$  & 3-10  & IRAS 1.9Jy$^z$ & 35-40  &  ``
\\* \hline \rule[-0.6ex]{0mm}{3.2ex}
4-10 & $1.5 \pm 0.5$  &  $4.4 \pm 3.7$  & 0.1-50  & IRAS 1.2Jy$^z$ & 
30-180  & Bouchet et al. 1993 
 \\* \hline \rule[-0.6ex]{0mm}{3.2ex}
5-10 & $2.8 \pm 0.1$  &  $6.9 \pm 0.7$  &   8-32 & IRAS 1.2Jy$^z$  & 
30-130 & Kim \& Strauss 1998
\\* \hline \rule[-0.6ex]{0mm}{3.2ex}
7 & $2.4 \pm 0.3$  &  $13 \pm 2$   & 2-10 & Perseus-Pisces$^z$  & 30 & 
Ghigna et.al 1996
\\* \hline \rule[-0.6ex]{0mm}{3.2ex}
 --- & $2.4 \pm 0.2$  &  ---   &  ---  & CfA   &  & Peebles 1980 (eq.[57.9])
\\* \hline \rule[-0.6ex]{0mm}{3.2ex}
6 & $2.0 \pm 0.3$  &  $6.3 \pm 1.6$   & 1-14 & CfA & 25- 50 & Fry \& 
Gazta\~naga 1994 
\\* \hline \rule[-0.6ex]{0mm}{3.2ex}
7 & $1.9 \pm 0.2$  &  $5.1 \pm 1.3$   & 2-16 & CfA$^z$& 25-50 &   ``
\\* \hline \rule[-0.6ex]{0mm}{3.2ex}
7 & $1.8 \pm 0.2$  &  $5.4 \pm 2.2$ & 1-12 & SSRS  & 25-50 &  ``
\\* \hline \rule[-0.6ex]{0mm}{3.2ex}
8 & $1.8 \pm 0.2$  &  $5.2 \pm 1.3$ & 2-11 & SSRS$^z$ & 25-50 &  ``
\\* \hline  \hline \rule[-0.6ex]{0mm}{3.2ex}
$5.6 \pm 0.4$    & $1.8 \pm 0.3$  &  $3.3 \pm 1.5$   &  2-8  & CfA50$^z$& 
25 &Gazta\~naga 1992 
\\* \hline \rule[-0.6ex]{0mm}{3.2ex}
$11.3 \pm 0.8$  & $1.7 \pm 0.2$  &  $2.5 \pm 1.1$   &  4-12  & CfA80$^z$ 
& 40   &  ``
\\* \hline \rule[-0.6ex]{0mm}{3.2ex}
$9.8 \pm 2.2$  & $1.7 \pm 0.5$  &  $3.0 \pm 4.3$   &  8-22  & CfA92$^z$ & 
50  &  ``
\\* \hline \rule[-0.6ex]{0mm}{3.2ex}
$6.7 \pm 0.5$ & $1.4 \pm 0.5$  &  $1.7 \pm 2.3$  &  2-8  & SSRS50$^z$ & 
25  & ``
\\* \hline \rule[-0.6ex]{0mm}{3.2ex}
$9.8 \pm 2.0$ & $1.9 \pm 0.4$  &  $4.4 \pm 2.0$    & 4-12  & SSRS80$^z$ & 
40 & ``
\\* \hline
$11.2 \pm 2.5$ & $2.2 \pm 0.9$  &  $6.6 \pm 8.5$    & 8-22  & SSRS115$^z$ 
& 60  & ``
\\* \hline
\end{tabular}

\caption[junk]{Some measurements of
the variance 
and $S_N$, for $N = 3, 4$, in the literature.
The first column $R_0$ is the scale at which the
measured variance equals $1$. The second and third
columns give $S_3$ and $S_4$, from either
counts-in-cells or the multi-point ratios $Q_N$,
at the scales specified in the fourth column.
In most cases, only the mean values for $S_N$ over a
range of scales were published. In cases where measurements of
the individual $S_N$ for each smoothing scale
are reported in the literature, we quote the actual
range of estimates over the corresponding range of scales.
An estimate of the effective radius 
of each sample is given by $R_L \equiv (\Omega/4\pi)^{1/3}\cal{D}$,
where $\Omega$ is the solid angle of the survey and $\cal{D}$ is taken
to be either the maximum depth in volume limited samples or twice the
mean depth in magnitude limited samples.
The effective volume for IRAS has been divided by two because there are two 
disconnected polar caps, and measurements were done by averaging
the results from the two pieces. Samples with the $z$-superscript
are in redshift-space, and those without are in angular-space.}
\label{s3s4}
\end{table}

\pagebreak
\clearpage
\newpage

\end{document}